\documentclass[twocolumn]{aastex63}

\newcommand\aastex{AAS\TeX}

\usepackage{savesym}
\savesymbol{tablenum}
\usepackage{siunitx}
\restoresymbol{SIX}{tablenum}

\usepackage[version=4]{mhchem}
\usepackage{graphicx,epsf}
\usepackage{subfigure}
\usepackage{graphicx}	
\usepackage{amsmath}	
\usepackage{amssymb}	
\usepackage{units}
\usepackage{newtxtext,newtxmath}
\usepackage{siunitx}
\usepackage{footnote}

\newcommand{\Msun}{{\rm M}_\odot}
\newcommand{\Rsun}{{\rm R}_\odot}

\newcommand{\kms}{\textrm{km}\,\textrm{s}^{-1}}
\newcommand{\niVI}{${}^{56}\textrm{Ni}$}
\newcommand{\coVI}{${}^{56}\textrm{Co}$}
\newcommand{\coVII}{${}^{57}\textrm{Co}$}

\def\ca{{CaST}}
\def\cas{{CaSTs}}

\shorttitle{SN 2019ehk Late-time}
\shortauthors{Jacobson-Gal\'an et al.}

\begin{document}

\title{Template \aastex Article with Examples: 
v6.3\footnote{Released on June, 10th, 2019}}

\title{Late-time Observations of Calcium-Rich Transient SN~2019ehk Reveal a Pure Radioactive Decay Power Source}

\correspondingauthor{Wynn Jacobson-Gal\'{a}n (he, him, his)}
\email{wynn@u.northwestern.edu}

\author[0000-0002-3934-2644]{Wynn V. Jacobson-Gal\'{a}n}
\affil{Department of Physics and Astronomy, Northwestern University, 2145 Sheridan Road, Evanston, IL 60208, USA}
\affil{Center for Interdisciplinary Exploration and Research in Astrophysics (CIERA), 1800 Sherman Ave, Evanston, IL 60201, USA}

\author[0000-0003-4768-7586]{Raffaella Margutti}
\affil{Department of Physics and Astronomy, Northwestern University, 2145 Sheridan Road, Evanston, IL 60208, USA}
\affil{Center for Interdisciplinary Exploration and Research in Astrophysics (CIERA), 1800 Sherman Ave, Evanston, IL 60201, USA}

\author{Charles~D.~Kilpatrick}
\affil{Department of Astronomy and Astrophysics, University of California, Santa Cruz, CA 95064,
USA}

\author[0000-0002-7868-1622]{John Raymond}
\affil{Center for Astrophysics \textbar{} Harvard \& Smithsonian, 60 Garden Street, Cambridge, MA 02138, USA}

\author{Edo Berger}
\affil{Center for Astrophysics \textbar{} Harvard \& Smithsonian, 60 Garden Street, Cambridge, MA  02138, USA}

\author[0000-0003-0526-2248]{Peter K. Blanchard}
\affil{Department of Physics and Astronomy, Northwestern University, 2145 Sheridan Road, Evanston, IL 60208, USA}
\affil{Center for Interdisciplinary Exploration and Research in Astrophysics (CIERA), 1800 Sherman Ave, Evanston, IL 60201, USA}

\author[0000-0002-4674-0704]{Alexey Bobrick}
\affil{Lund University, Department of Astronomy and Theoretical physics, Box 43, SE 221-00 Lund, Sweden}

\author{Ryan J. Foley}
\affil{Department of Astronomy and Astrophysics, University of California, Santa Cruz, CA 95064,
USA}

\author[0000-0001-6395-6702]{Sebastian Gomez}
\affil{Center for Astrophysics \textbar{} Harvard \& Smithsonian, 60 Garden Street, Cambridge, MA 02138, USA}

\author[0000-0002-0832-2974]{Griffin Hosseinzadeh}
\affil{Center for Astrophysics \textbar{} Harvard \& Smithsonian, 60 Garden Street, Cambridge, MA 02138, USA}

\author[0000-0002-0763-3885]{Danny Milisavljevic}
\affil{Department of Physics and Astronomy, Purdue University, 525 Northwestern Avenue, West Lafayette, IN 47907, USA}

\author{Hagai Perets}
\affil{Technion - Israel Institute of Technology, Physics department, Haifa Israel 3200002}

\author{Giacomo Terreran}
\affil{Department of Physics and Astronomy, Northwestern University, 2145 Sheridan Road, Evanston, IL 60208, USA}
\affil{Center for Interdisciplinary Exploration and Research in Astrophysics (CIERA), 1800 Sherman Ave, Evanston, IL 60201, USA}

\author{Yossef Zenati}
\affil{Technion - Israel Institute of Technology, Physics department, Haifa Israel 3200002}

\begin{abstract}

We present multi-band \textit{Hubble Space Telescope} imaging of the Calcium-rich supernova (SN) 2019ehk at 276 - 389~days after explosion. These observations represent the latest $B$-band to NIR photometric measurements of a Calcium-rich transient to date and allow for the first opportunity to analyze the late-time bolometric evolution of an object in this observational SN class. We find that the late-time bolometric light curve of SN~2019ehk can be described predominantly through the radioactive decay of \coVI \ for which we derive a mass of $M({}^{56}\textrm{Co}) = (2.8 \pm 0.1) \times 10^{-2}$~$\Msun$. Furthermore, the rate of decline in bolometric luminosity requires the leakage of $\gamma$-rays on timescale $t_{\gamma} = 53.9 \pm 1.30$~days, but we find no statistical evidence for incomplete positron trapping in the SN ejecta. While our observations cannot constrain the exact masses of other radioactive isotopes synthesized in SN~2019ehk, we estimate a mass ratio limit of $M({}^{57}\textrm{Co}) / M({}^{56}\textrm{Co}) \leq 0.030$. This limit is consistent with the explosive nucleosynthesis produced in the merger of low-mass white dwarfs, which is one of the favored progenitor scenarios in early-time studies of SN~2019ehk. 

\end{abstract}


\section{Introduction} \label{sec:intro}

Calcium-rich (Ca-rich) transients are a peculiar class of thermonuclear transients that were identified almost two decades ago and studied extensively ever since \citep{filippenko03, perets10, kasliwal12}. These supernovae (SNe) are defined observationally by fast-evolving light curves ($t_r < 15$~days) and low overall luminosities ($M_{peak} > -16$~mag), both photometric properties being consistent with typical ejecta and \niVI \ mass estimates of $\lesssim 0.5~\Msun$ and $\lesssim 0.1~\Msun$, respectively \citep{taubenberger17}. The ``Ca-rich'' naming convention is in part derived from their spectroscopic evolution wherein these transients exhibit prominent [\ion{Ca}{ii}] emission features in their photospheric and nebular phase spectra compared to [\ion{O}{i}] emission ([\ion{Ca}{ii}]/[\ion{O}{i}] > 2; \citealt{milisavljevic17}). However, while Ca-rich transients appear to cool most efficiently through \ion{Ca}{ii} transitions over \ion{O}{i}, it is debated whether these explosions are in fact more abundant in calcium ions than oxygen by mass \citep{perets10, milisavljevic17}. As a result, we choose to adopt the nomenclature presented by \cite{shen19} and refer to these objects as ``Calcium-strong transients'' (\cas) throughout the paper. 

The majority of known \cas \ are located in the outskirts of early-type host galaxies \citep{perets11, kasliwal12}. However, as the number of confirmed \cas \ increases, there appears to be a substantial spread in their host morphology that includes both disk-shaped as well as elliptical galaxies \citep{perets10,Perets2014, milisavljevic17, de20}. Additionally, \cas \ are typically found in galaxy groups or cluster environments with no evidence of star formation and their explosion sites are generally associated with old stellar populations \citep{perets10, perets11, lyman14, foley15,lunnan17}. Consequently, typical progenitor systems proposed for \cas \ have included a white dwarf (WD) with a neutron star (NS), a black hole (BH), another WD or a non-degenerate main sequence star companion \citep{rosswog08, perets10, metzger12, macleod14, sell15, margalit16,Bobrick2017,Zenati2019, zenati2019b}. Nevertheless, the observed diversity in host galaxies and explosion characteristics suggests heterogeneity amongst \ca \ progenitors \citep{milisavljevic17}. Therefore, increasing the sample size of objects and performing novel studies of new \cas \ will help uncover the origins of this unique explosion class.  

On 2019 April 29 (MJD 58602.24), the closest known \ca, SN~2019ehk, was detected in the nearby galaxy NGC~4321 (M100) at $16.2 \pm 0.4 $~Mpc \citep{wjg20, Nakaoka20}. Observations of SN~2019ehk were acquired as early as $\sim$10 hours after explosion ($t_{exp} = 58601.8 \pm 0.1$~days, in MJD), which allowed for unprecedented multi-wavelength coverage of this event. Fast-cadence observations revealed a double-peaked light curve in optical bands, with the primary peak being temporally coincident with luminous X-ray emission ($L_X\approx10^{41}~\rm{erg~s^{-1}}$); the first instance of X-ray detections in a \ca. Combined with flash-ionized H$\alpha$ and \ion{He}{ii} spectral lines at +1.5d after explosion, these observations revealed the presence of dense circumstellar material (CSM) in a compact shell surrounding the progenitor system at the time of explosion. \cite{wjg20} (hereafter WJG20a) also presented deep \textit{Hubble Space Telescope (HST)} pre-explosion imaging of the explosion site that constrained the possible progenitor of SN~2019ehk to be either a massive star in the lowest mass bin ($\lesssim$~10~$\Msun$) or a WD in a binary system. Alternatively, \cite{Nakaoka20} suggest that SN~2019ehk is an ultra-stripped SN candidate that arose from a He (or C/O) star + NS binary configuration. This latter scenario, however, is difficult to reconcile with the presence of H-rich material in the local circumstellar environment. Recently, based on derived Oxygen mass, \cite{de20b} concluded that the progenitor of SN~2019ehk was a low-mass massive star ($M_{\rm ZAMS} \approx 9-9.5~\Msun$) that lost most of its H envelope via binary interaction prior to explosion. We explicitly address the viability of this alternative scenario of a ``Calcium-rich type IIb'' SN proposed by \cite{de20b} in Section \ref{subsec:Omass} and we offer an additional, independent calculation of the Oxygen mass parameter.

Photometric observations of SNe at late time phases ($t \gtrsim 300$~days) enables the study of explosion power sources and, consequently, the progenitor system responsible for a given transient. To date, only a few \cas \ and \ca \ candidates have been detected in photometric observations at $\gtrsim$~250~days after explosion e.g., PTF10iuv \citep{kasliwal12}, SN~2012hn \citep{valenti14}, and SN~2018gwo \citep{de20}. The close proximity of SN~2019ehk provides the first opportunity to accurately reconstruct the late-time bolometric light-curve evolution of a CaST using multi-color observations that span from $B$-band to the NIR at $t \gtrsim 250$~days. In this paper, we present late-time \textit{HST} observations of SN~2019ehk and modeling of the bolometric light curve out to $\sim$~400\,d post-explosion. In \S\ref{sec:obs} we present observations and data reduction of SN~2019ehk. In \S\ref{sec:analysis} we present modeling of SN~2019ehk's bolometric light curve evolution and derive physical properties of the radioactive decay-powered explosion. In \S\ref{sec:discussion} we discuss how SN~2019ehk compares to other late-time SN light curves and how these new observations constrain the SN progenitor system. 

\section{Observations} \label{sec:obs}

Early-time observations of SN~2019ehk were conducted with a variety of ground-based telescopes from 28 April to 02 August 2019 ($\sim$~$0.5-96.2$~days after explosion). Specifics about reductions techniques and instruments used are presented in WJG20a. Following WJG20a, we adopt a host galaxy distance of $16.2 \pm 0.400 $~Mpc, distance modulus $\mu = 31.1 \pm 0.100$~mag, redshift $z = 0.005 \pm 0.0001$ and time of explosion $t_{exp} = 58601.8 \pm 0.1$~days (MJD). The Milky Way color excess along the SN line of sight is \textit{E(B-V)} = 0.0227~mag \citep{schlegel98, schlafly11} and the host galaxy reddening is $E(B-V)= 0.47 \pm 0.10$~mag\footnote{\citealt{Nakaoka20} and \citealt{de20b} assume a host galaxy reddening range of 0.5-1.0~mag that is derived from a comparison between SN~2019ehk and two particular SNe. Our adopted color excess from WJG20a lies at the lower end of this range and it is based on (i) direct measurements of Balmer decrement of the \ion{H}{ii} region from pre-explosion spectroscopy of the SN explosion site and (ii) color comparisons to \ca \ and SNe~Ic samples (e.g., Fig. 10 in WJG20a).} (WJG20a), both of which we correct for using a standard \cite{fitzpatrick99} reddening law (\textit{$R_V$} = 3.1). Understanding if alternative \textit{$R_V$} values are more appropriate descriptors of the host galaxy extinction is beyond the scope of this paper and we proceed with \textit{$R_V$} = 3.1 so as to remain consistent with other studies on SN~2019ehk.

We obtained additional late-time, ground-based imaging of SN~2019ehk on 1 January 2020 ($\sim$247.2~days after explosion) in $r-$ and $i-$band with the Inamori-Magellan Areal Camera and Spectrograph (IMACS; \citealt{dressler11}) on the Magellan Baade 6.5-m Telescope. The data were first bias-subtracted and flat-fielded, then three frames per filter were averaged using \texttt{PyRAF}. From these observations, we measure an $i$-band AB magnitude of $21.40 \pm 0.06$~mag and derive an $r$-band upper limit of $>$~23.51~mag.

\begin{figure*}[t]
\centering
\includegraphics[width=0.95\textwidth]{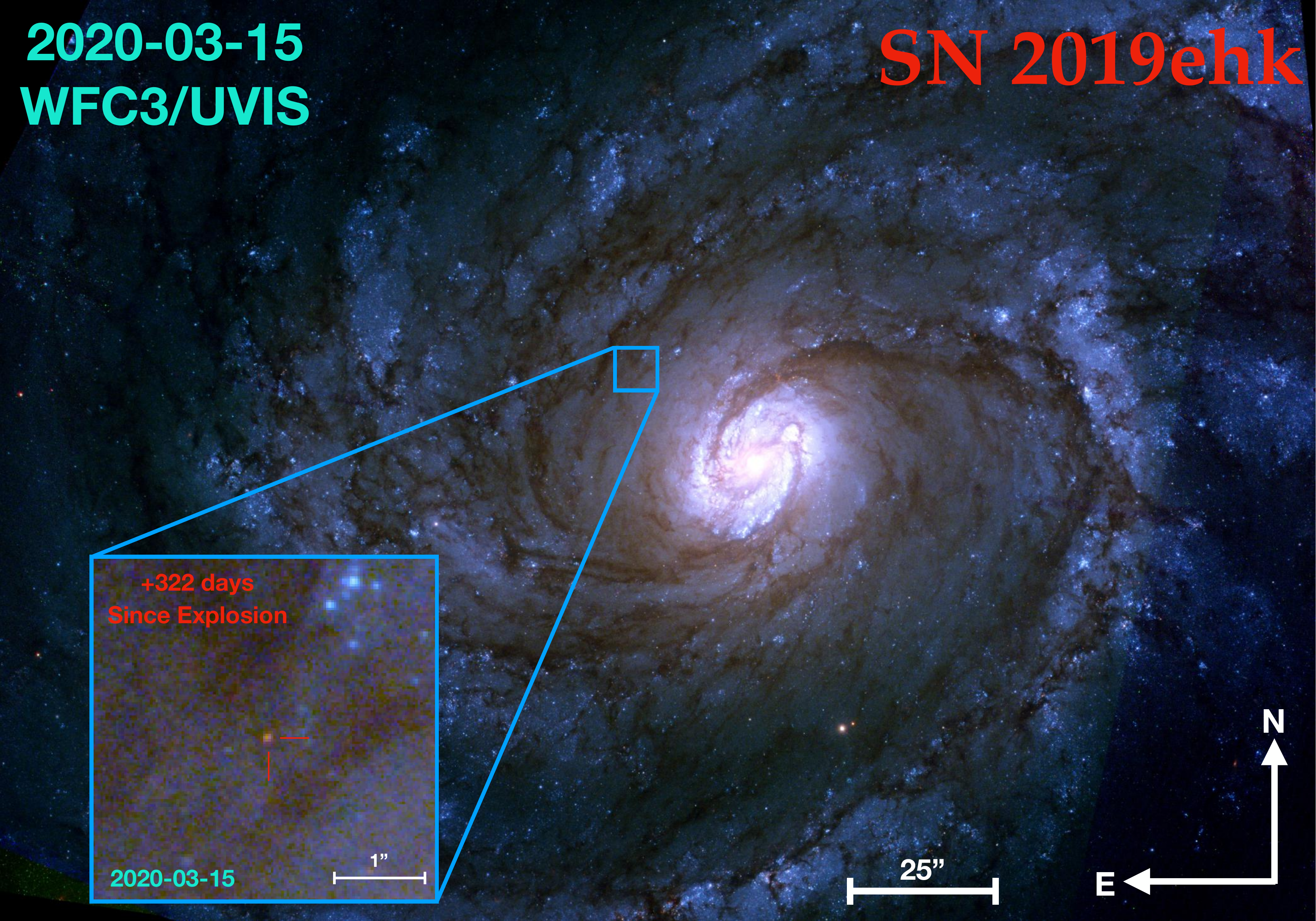} 
\caption{False color, \textit{HST} RGB  image of SN~2019ehk in host galaxy M100 at +322 days after explosion. The SN is marked in red within the zoomed-in blue box in lower left corner.  \label{fig:hst}}
\end{figure*}

Late-time \textit{HST} imaging of SN~2019ehk was first obtained in F275W, F336W, F438W, F555W and F814W filters (2000-10000\AA) with the Wide Field Camera 3 (WFC3) through \textit{HST} program PID-15654 (PI Lee) on 29 January and 15 March 2020 ($\sim$276.2 and 321.8d after explosion, respectively). Additional UVIS/IR WFC3 imaging was taken in F555W, F814W, F110W and F160W filters (0.45 - 1.7~$\mu$m) on 21 May 2020 ($\sim$389.0d after explosion) under \textit{HST} program PID-16075 (PI Jacobson-Gal\'an). Following methods in \cite{kilpatrick18}, we reduced all {\it HST} imaging using the {\tt hst123}\footnote{\url{https://github.com/charliekilpatrick/hst123}} python-based reduction package.  We downloaded all relevant calibrated WFC3/UVIS and IR images ({\tt flc}/{\tt flt} frames) from the Mikulski Archive for Space Telescopes.\footnote{\url{https://archive.stsci.edu/hst/}}  Each image was then aligned to a common reference frame using {\tt TweakReg}.  We then drizzled images from common filters and epochs using {\tt astrodrizzle}.  Finally, we performed photometry in the original, calibrated images using {\tt dolphot} \citep{dolphot}. We present the observed apparent magnitudes (AB system), as well as 3-$\sigma$ upper limits derived from fake star injection, for all late-time \textit{HST} filters in Table \ref{tbl:hst_table}. The late-time false color RGB image of SN~2019ehk and its host galaxy is shown in Figure \ref{fig:hst}.  The complete multi-band light curve of SN~2019ehk is shown in Figure \ref{fig:LCs}(a). 

\section{Analysis} \label{sec:analysis}

In \S\ref{subsec:bol_lc} we describe the derivation of SN~2019ehk's bolometric light curve which spans $\sim$0.5 to 388.8~days after explosion. In \S\ref{subsec:modeling} we apply an analytic formalism for a radioactive-decay powered emission to fit the late-time light curve evolution of SN~2019ehk and derive physical parameters of the explosion. 

\subsection{Pseudo-Bolometric Light Curve} \label{subsec:bol_lc}
At $t<97$~days, we construct a pseudo-bolometric light curve of SN~2019ehk through a combination of multi-band photometry from multiple ground-based telescopes (e.g., see WJG20a). For each epoch, luminosities are calculated through trapezoidal integration of SN flux in $uBVcgoriz$ bands (3000-10000\,\AA). Uncertainties are estimated through a Monte Carlo simulation that includes 1000 realizations of the data. In time intervals without complete color information, we interpolated between light curve data points using a low-order polynomial spline. This method is different than that used by WJG20a who created a bolometric light curve of SN~2019ehk through fitting of the spectral energy distribution (SED) with a blackbody model. The two methods lead to consistent luminosities for $t \lesssim 40$~days. However, the blackbody model over-predicts the total flux at later phases due to the prominent [\ion{Ca}{ii}] and \ion{Ca}{ii} line transitions that dominate the SED flux in some bands. As expected, the blackbody model becomes an inadequate description of the observed emission as soon as the SN transitions to an emission dominated spectrum in the nebular phase. Therefore, we apply the trapezoidal integration method to determine the bolometric luminosity at all phases for consistency. 

\begin{figure*}[t]
\centering
\subfigure[]{\includegraphics[width=0.58\textwidth]{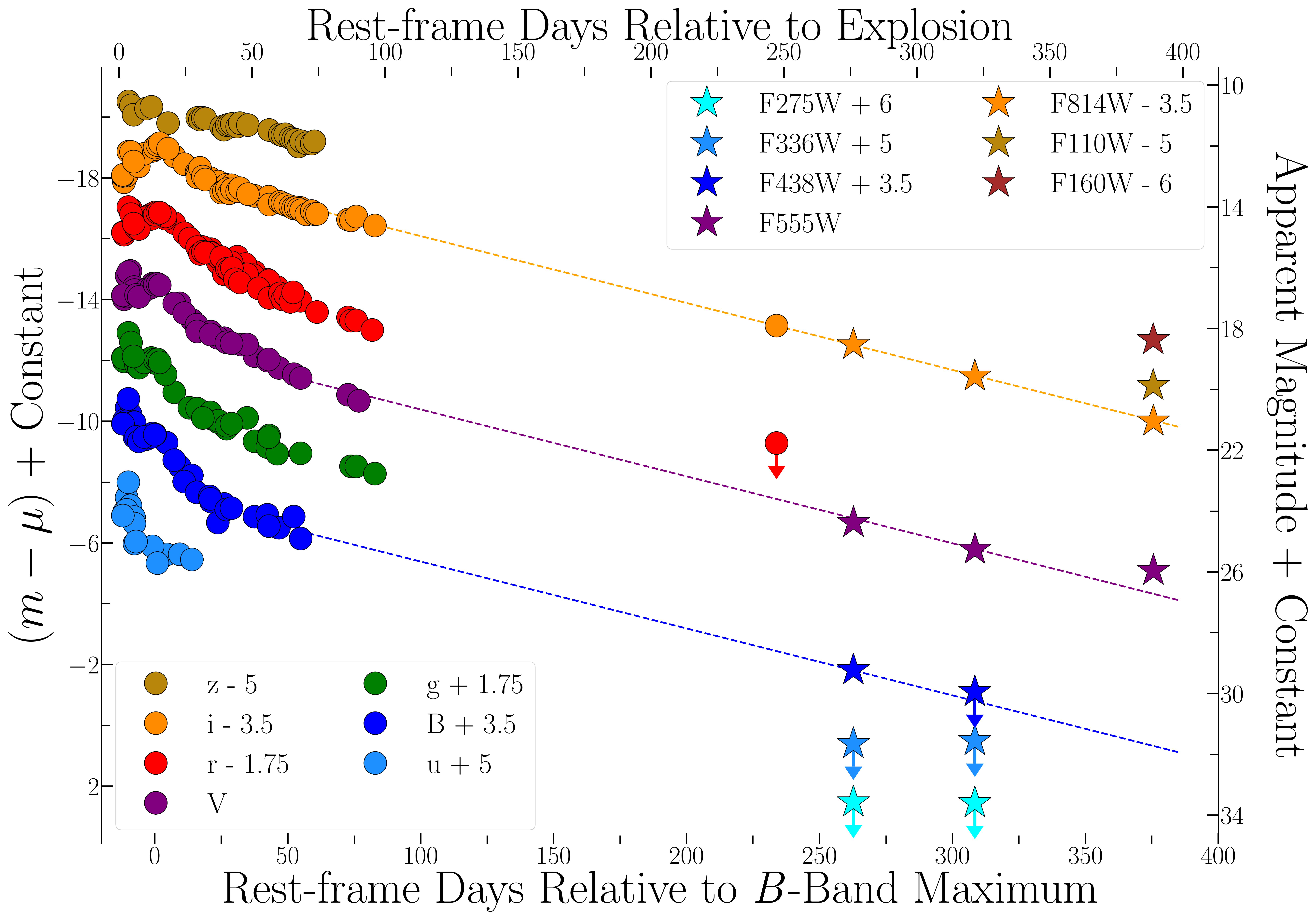}}
\subfigure[]{\includegraphics[width=0.41\textwidth]{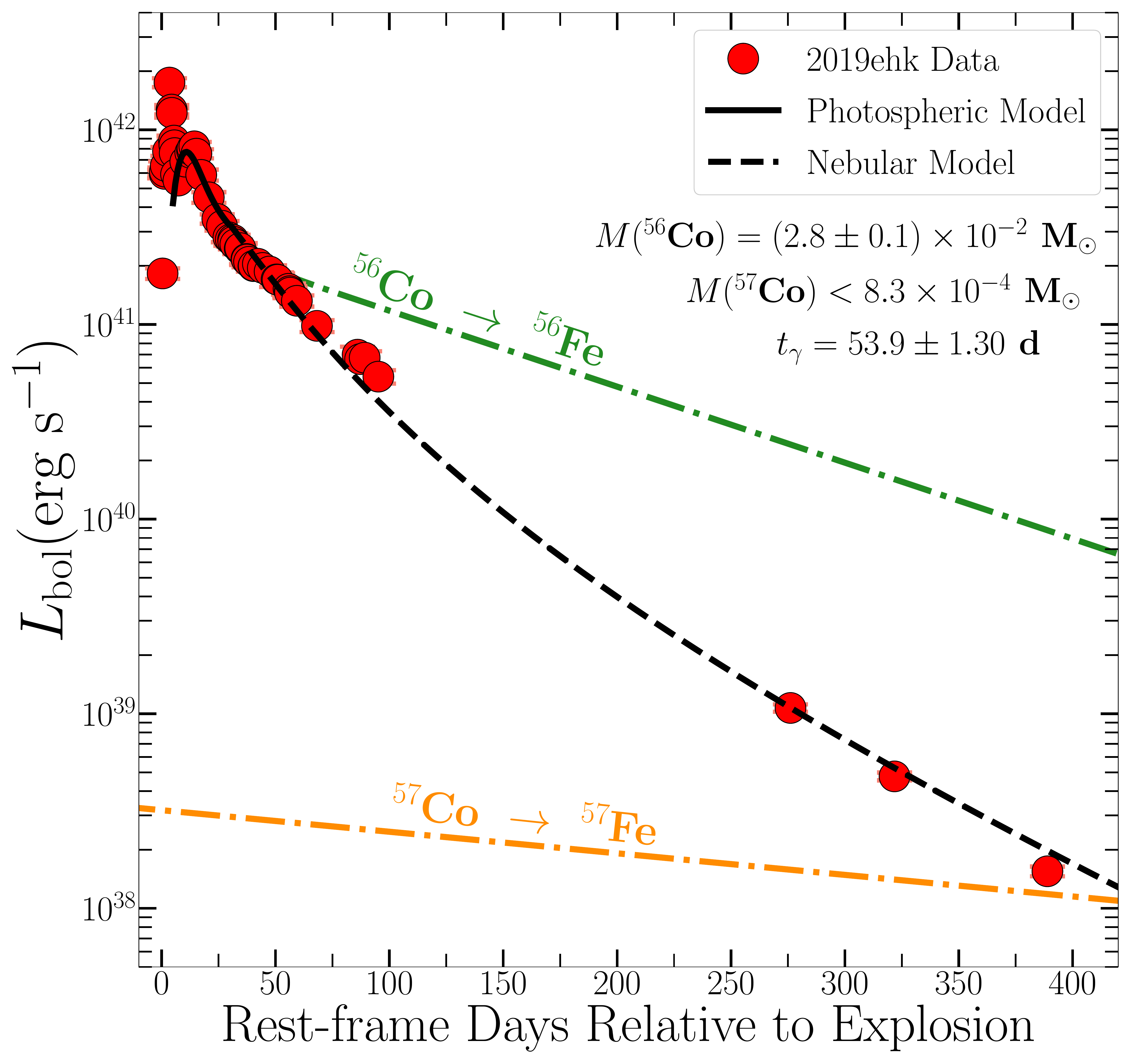}}
\caption{(a) UV/Optical/IR light curve of SN~2019ehk with respect to second $B$-band maximum. Observed photometry is presented in the AB magnitude system. Circles denote ground-based photometry, the majority of which were presented in WJG20a. Stars represent late-time \textit{HST} detections and upper limits. Dashed lines provide visual extrapolation between early and late-time data in filters that are roughly consistent between \textit{HST} and ground-based filters. (b) Bolometric light curve data of SN~2019ehk shown in red circles with respect to the radioactive decay model fit (dashed line) discussed in \S\ref{subsec:modeling}. Solid black line is the photospheric model from WJG20a that fits the early-time data around peak using SN parameters $M({}^{56}\textrm{Ni})$, $E_k$ and $M_{ej}$. Dotted-dashed green and orange lines represent the decays of \coVI \ and \coVII, respectively, for the complete trapping of $\gamma$-rays and positrons. \label{fig:LCs} }
\end{figure*}

For late-time observations at $t > 276$~days, we also perform trapezoidal integration of SN~2019ehk's spectral energy distribution (SED) in \textit{HST} filters (0.3-1.7~$\mu$m). Because infrared (IR) \textit{HST} imaging was only taken during the last epoch (+389d, Fig. \ref{fig:LCs}a), we extrapolate backwards in time in order to apply an IR correction that constitutes $\sim$30\% of the bolometric flux to the \textit{HST} observations at +276 and +322d after explosion. We proceed with the assumption that such a correction is not necessary for early-time epochs ($t \lesssim 100$~days) where IR contribution is negligible. Furthermore, we note that there may be a small fraction ($\lesssim$~5\%) of UV SN flux that is not taken into account when constructing the late-time bolometric luminosities due to observed non-detections in the F275W, F336W and F475W \textit{HST} filters.  The complete bolometric light curve of SN~2019ehk is presented in Table \ref{tbl:bol_table} and plotted in Figure \ref{fig:LCs}(b). 

\subsection{Radioactive Decay Model} \label{subsec:modeling}

The late-time light curve evolution of most SNe is governed primarily by the energy deposition rate of the radioactive decay chain $\ce{{}^56Ni ->[ t_{\rm decay} = 8.77 \textrm{d} ] {}^56Co ->[ t_{\rm decay} = 111.3 \textrm{d} ] {}^56Fe }$ \citep{arnett82}. $\gamma$-rays released in this process are then thermalized in the expanding SN ejecta and, for phases $t \gtrsim 60$~days after explosion, \coVI \ beta-decay will power the bolometric light curve until the decays of other radioactive species such as \coVII \ and ${}^{55}\textrm{Fe}$ become dominant (e.g., $t \gtrsim 500$~days after explosion; orange dot-dashed line in Fig. \ref{fig:LCs}b). 

In this section, we describe the components of a purely radioactive decay powered model and apply it to fit the bolometric light curve of SN~2019ehk at late times. The total energy generated in each beta-decay can be defined by (i) $\gamma$-rays released in the decay chain, (ii) kinetic energy of emitted positrons and (iii) $\gamma$-rays produced from electron-positron annihilation. Regardless of the generation process, all $\gamma$-rays produced have a finite probability of escaping the ejecta before depositing their energy. The limiting case where the $\gamma$-ray photons from the \coVI \  decay are completely trapped and thermalized within the expanding ejecta is shown in Fig. \ref{fig:LCs}b (green dot-dashed line). However, observations of Type Ia SNe (SNe Ia) and stripped envelope Type Ib/c SNe (SNe Ib/c) clearly show more rapid light curve decays, indicating that a fraction of $\gamma$-rays is able to escape before depositing their energy into the ejecta \citep{Cappellaro97, wheeler15}. Following \cite{clocchiatti97}, the $\gamma$-ray leakage can be parameterized in terms of a trapping timescale, $t_{\gamma}$. The kinetic energy from positrons can also be thermalized and therefore their potential leakage from the SN ejecta can be described by a positron trapping timescale, $t_{e^+}$.

To model the late-time light curve of SN~2019ehk, we apply the formalism outlined in the appendix of \cite{valenti08} for energy deposition from radioactive decay during the nebular phase ($t \gtrsim 60$~days). This model is very similar to the Bateman equation (e.g., see eqn. 16 in \citealt{seitenzahl14}) in how it can be used to derive masses of radioactive isotopes and trapping timescales, $t_{\gamma}$ and $t_{e^+}$. However, unlike the Bateman formalism, this method self-consistently accounts for the trapping of $\gamma$-rays created through electron-positron annihilation.  

The total luminosity generated by radioactive decay of \niVI \ and ${}^{57}\textrm{Ni}$ during the nebular phase ($t \gtrsim 60$~days) is described by the following expression, originally presented by \cite{sutherland84} and \cite{Cappellaro97} and summarized here for clarity:  

\begin{equation}\label{eq:Lneb}
    \begin{aligned}
    L_{\rm neb} (t) = S^{{}^{56}\textrm{Ni}}(\gamma) \ + \ S^{{}^{56}\textrm{Co}}(\gamma) \ + \  S^{{}^{56}\textrm{Co}}_{e^+} (\gamma) \ +\\  S^{{}^{56}\textrm{Co}}_{e^+} (KE) \ + \  S^{{}^{57}\textrm{Co}}(\gamma)
    \end{aligned}
\end{equation}

$S^{{}^{56}\textrm{Ni}}(\gamma)$ is the energy deposition due to \niVI \ decay:

\begin{equation}\label{eq:SNi}
    S^{{}^{56}\textrm{Ni}}(\gamma) = M({}^{56}\textrm{Ni})\epsilon_{{}^{56}\textrm{Ni}}e^{-t/\tau_{{}^{56}\textrm{Ni}}}
\end{equation}

where $M({}^{56}\rm{Ni})$ is the mass of $^{56}$Ni, $\epsilon_{{}^{56}\textrm{Ni}} = 3.9 \times 10^{10}$~erg s$^{-1}$ g$^{-1}$ is the energy rate generated by the decay of $^{56}$Ni per unit mass and a decay time scale of $\tau_{{}^{56}\textrm{Ni}} = 8.77$~days. The remaining terms in Equation \ref{eq:Lneb} constitute the energy deposition rate due to the respective decays of \coVI \ and \coVII. 81\% of the energy  released by the \coVI \ decay is emitted in the form of $\gamma$-rays:

\begin{equation}\label{eq:SCo56}
    S^{{}^{56}\textrm{Co}}(\gamma) = 0.81\mathcal{E} \Big(1 - e^{-(t_{\gamma}/t)^2} \Big)
\end{equation}

The term $(1 - e^{-(t_{\gamma}/t)^2})$ accounts for the trapping probability of the $\gamma$-rays in the ejecta and $\mathcal{E}$ is the rate of energy production from the $^{56}$Co decay:
    
\begin{equation}\label{eq:epsilon}
    \mathcal{E} = M({}^{56}\textrm{Ni})\epsilon_{{}^{56}\textrm{Co}} \Big ( e^{-t/\tau_{{}^{56}\textrm{Co}}} -  e^{-t/\tau_{{}^{56}\textrm{Ni}}} \Big )
\end{equation}

where $\epsilon_{{}^{56}\textrm{Co}} = 6.8 \times 10^{9}$~erg s$^{-1}$ g$^{-1}$ and $\tau_{{}^{56}\textrm{Co}} = 111.3$~days. The remaining 19\% of energy from \coVI \ decay is released via positrons and is partly described by the following expression for energy deposition from $\gamma$-rays created in positron annihilation: 

\begin{equation}\label{eq:SCo56_positron}
    S^{{}^{56}\textrm{Co}}_{e^+} (\gamma) = 0.164\mathcal{E} \Big(1 - e^{-(t_{\gamma}/t)^2} \Big)\Big(1 - e^{-(t_{e^+}/t)^2} \Big)
\end{equation}

where the terms $(1 - e^{-(t_{\gamma}/t)^2})$ and $(1 - e^{-(t_{e^+}/t)^2})$ account for the trapping probabilities of the $\gamma$-rays and positrons, respectively. The remaining source of energy in \coVI \ decay is kinetic energy from positrons and is expressed by:

\begin{equation}\label{eq:SCo56_KE}
    S^{{}^{56}\textrm{Co}}_{e^+} (KE) = 0.036\mathcal{E} \Big(1 - e^{-(t_{e^+}/t)^2} \Big)
\end{equation}

Lastly, we consider the contribution of \coVII \ decay to the bolometric light curve, which we parameterize as follows: 

\begin{equation}\label{eq:SCo57}
    S^{{}^{57}\textrm{Co}}(\gamma) = M({}^{57}\textrm{Co})\epsilon_{{}^{57}\textrm{Ni}}e^{-t/\tau_{{}^{57}\textrm{Co}}}
\end{equation}

where $\epsilon_{{}^{57}\textrm{Ni}} = 8.9 \times 10^{6}$~erg s$^{-1}$ g$^{-1}$ for no $\gamma$-ray trapping, $\epsilon_{{}^{57}\textrm{Ni}} = 7.0 \times 10^{7}$~erg s$^{-1}$ g$^{-1}$ for complete $\gamma$-ray trapping, and $\tau_{{}^{57}\textrm{Co}} = 392.11$~days. We adopt the energy rate $\epsilon_{{}^{57}\textrm{Ni}}$ that assumes no trapping of $\gamma$-rays and complete trapping of X-rays and Auger electrons (e.g., see \citealt{seitenzahl2009}, \citealt{Graur16}). This description of energy deposition from $\gamma$-rays released in \coVII \ decay will yield the most conservative estimate on the total \coVII \ mass in SN~2019ehk. We also ignore any ``freeze-out'' effects that typically influence the SN light curve at $t > 600$~days \citep{frannson93, Fransson15, graur17}. 

In this model, free variables include the total masses of \coVI \ and \coVII \ as well as the timescales of $\gamma$-ray and positron escape, $t_{\gamma}$ and $t_{e^{+}}$, respectively. We do not fit for other physical parameters that define these timescales such as the density profile, opacity, mass and kinetic energy of the expanding ejecta. These dependencies are discussed in the context of derived trapping timescales in Equations \ref{eq:F} and \ref{eq:C}. To fit the bolometric light curve, we use the non-linear least squares package \texttt{curve\_fit} in \texttt{scipy} \citep{virtanen20}. Our final model fit to the late-time light curve is shown as the dashed black line in Figure \ref{fig:LCs}(b). 

Using Equation \ref{eq:Lneb}, we first attempt to model the bolometric light curve of SN~2019ehk with the inclusion of partial trapping of positrons e.g., including $t_{e^+}$. We find that the model is insensitive to the positron trapping timescale and no meaningful statistical boundary can be constrained. We then model the bolometric light curve under the assumption of complete positron trapping (i.e., $(1 - e^{-(t_{e^+}/t)^2}) = 1$) and derive a total \coVI \ mass of $M({}^{56}\textrm{Co}) = (2.8 \pm 0.10) \times 10^{-2}$~$\Msun$ and a $\gamma$-ray trapping timescale of $t_{\gamma} = 53.9 \pm 1.30$~days. The estimated \coVI \ mass is consistent with the \niVI \ mass of $M_{\textrm{Ni}}=(3.1 \pm 0.11) \times 10^{-2} \ \Msun$ derived from photospheric modeling of the SN~2019ehk light curve at $t < 30$~days after explosion (WJG20a). This indicates that the early-time luminosity of SN~2019ehk during its second light curve peak was primarily powered by radioactive decay and not by additional power sources such as CSM interaction. Conversely, the first light curve peak at $t < 7$~days after explosion was powered by interaction with dense CSM (WJG20a).  

Because SN~2019ehk's bolometric light curve only extends to $\sim$400 days after explosion, the model fit is not fully sensitive to the contribution of \coVII \ decay to the overall SN luminosity. Consequently, we derive an upper limit on the total mass of \coVII \ in SN~2019ehk to be $M({}^{57}\textrm{Co}) < 8.3\times 10^{-4}$~$\Msun$, which represents a 3$\sigma$ statistical deviation relative to the late-time light curve data. Based on these mass estimates, we find a ratio of radioactive isotope masses in SN~2019ehk to be $M({}^{57}\textrm{Co}) / M({}^{56}\textrm{Co}) \leq 0.030$. As previously stated, this mass ratio represents the most conservative limit under the assumption of no $\gamma$-ray trapping from \coVII \ decay. However, for complete $\gamma$-ray trapping from this decay chain, the least conservative limit on \coVII \ mass in SN~2019ehk is $M({}^{57}\textrm{Co}) < 1.1\times 10^{-4}$~$\Msun$, which yields a mass ratio of $M({}^{57}\textrm{Co}) / M({}^{56}\textrm{Co}) \leq 0.004$. It is likely that the true \coVII \ mass and mass ratio limits for SN~2019ehk are within this range given the evidence of partial $\gamma$-ray trapping from \coVI \ decay at late-times. Finally, given the uncertainty on SN~2019ehk's host extinction, we also calculate the $M({}^{57}\textrm{Co}) / M({}^{56}\textrm{Co})$ mass ratio limit after correcting the data for a maximum color excess of $E(B-V) = 1$~mag as presented in the range by \cite{Nakaoka20}. We find mass ratio limits of $\leq 0.0044$ and $\leq 0.034$ for complete and no $\gamma$-ray trapping from \coVII \ decay, respectively. These limits are consistent to those calculated with our preferred host extinction value presented in \S\ref{sec:obs}. We also note that the estimated mass ratio limits are marginally dependent on the bolometric correction to the IR flux at late times as discussed in \S\ref{subsec:bol_lc}.

As shown by \cite{clocchiatti97}, the trapping timescales of both $\gamma$-rays and positrons are physical parameters that are defined based on properties of the SN ejecta. For $\gamma$-ray trapping, the expression is:

\begin{equation}\label{eq:F}
    t_{\gamma} = \Big( C(\eta) \kappa_{\gamma} M_{ej}^2 E_k ^{-1} \Big )^{1/2}
\end{equation}

where the ejecta opacity to $\gamma$~rays is $\kappa_{\gamma} = 0.027$~cm$^2$~g$^{-1}$, $M_{ej}$ is the ejecta mass, $E_k$ is the kinetic energy of the ejecta and the density function $C(\eta)$, under the assumption of spherical symmetry, is written as:

\begin{equation}\label{eq:C}
    C(\eta) = (\eta-3)^2 \Big [8\pi (\eta -1)(\eta-5) \Big]^{-1}
\end{equation}

where $\eta$ defines the density profile of ejecta i.e., $\rho_{ej} \propto r^{-\eta}$. 

Following \cite{valenti08}, we assume the ejecta is homogeneous and has a flat density profile of $\eta = 0$ within Eqn. \ref{eq:C}, which then yields $C(0) = 0.072$. For the known $\gamma$-ray energies of the beta-decays, the $\gamma$-ray opacity of the ejecta is expected to be $\kappa_{\gamma} = 0.027$~cm$^2$~g$^{-1}$ \citep{colgate80, woosley89, clocchiatti97}. To check this assumption, we solve for $\kappa_{\gamma}$ in Eqn. \ref{eq:F} using $C(0)$ listed above, $t_{\gamma}$ from our model fits as well as $M_{ej} = 0.7$~$\Msun$ and $E_k = 1.6 \times 10^{50}$~erg as derived in WJG20a from early-time light curve modeling. With these values, we estimate a $\gamma$-ray opacity of $\kappa_{\gamma} = 0.026 \pm 0.0019$~cm$^2$~g$^{-1}$, which is consistent with the fiducial value used in other studies. Furthermore, this agreement suggests that the SN 2019ehk ejecta structure can be consistent with being homogeneous and spherically symmetric, with synthesized Ni located at the center.

\begin{figure*}
\centering
\subfigure[]{\includegraphics[width=0.49\textwidth]{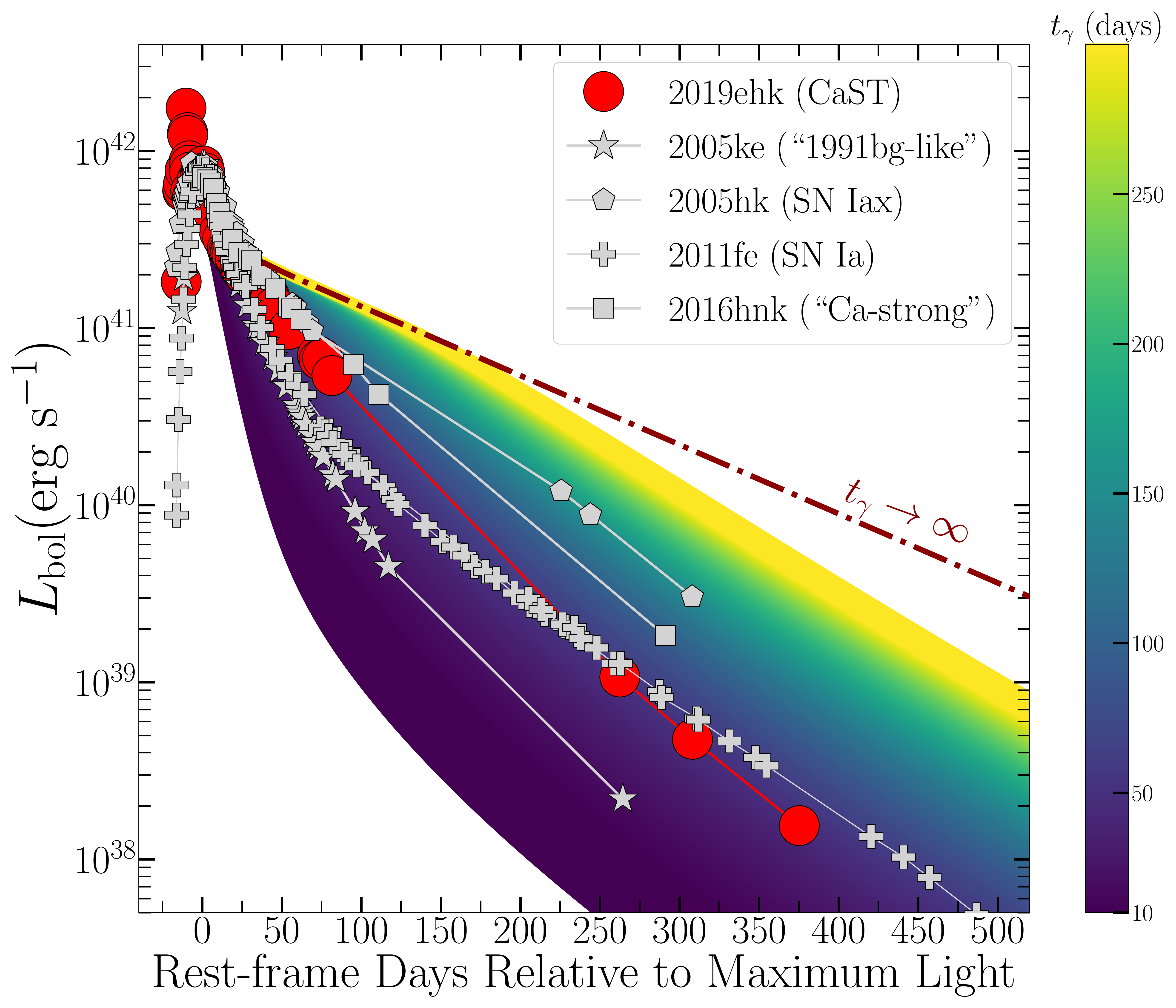}}
\subfigure[]{\includegraphics[width=0.49\textwidth]{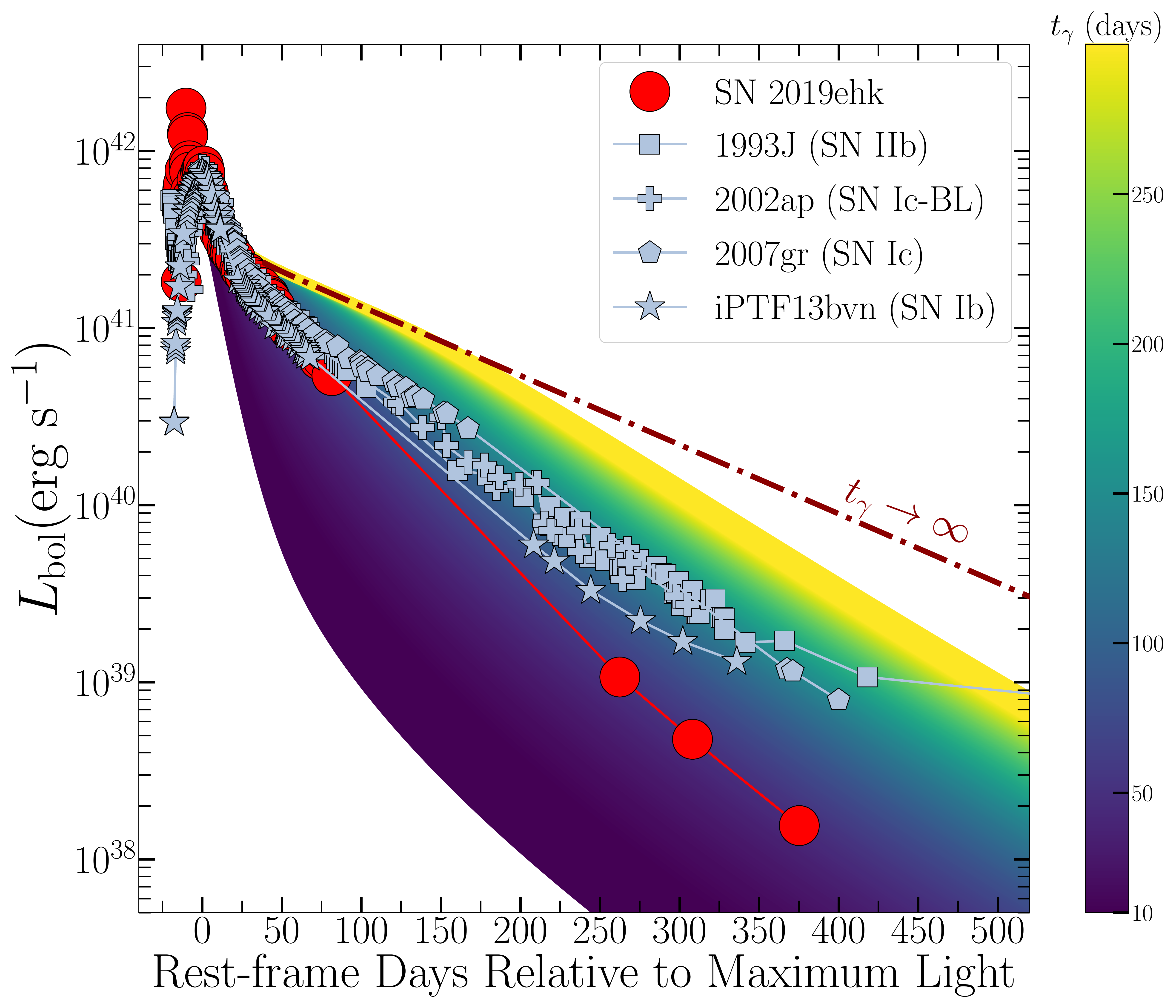}}
\caption{(a) Bolometric light curve comparison of SN~2019ehk and thermonuclear SN varieties such as SNe~Ia (SN~2011fe; grey plus signs, \citealt{zhang16}), 1991bg-like (SN~2005ke; grey stars, \citealt{contreras10}), SNe~Iax (SN~2005hk; grey polygons, \citealt{sahu08}) and Calcium-strong SNe (SN~2016hnk; grey squares, \citealt{galbany19, wjg19}). Comparison objects have been shifted in luminosity and phase space to match SN~2019ehk's light curve maximum. Dark red dashed-dotted line represents a radioactive decay light curve model fit to SN~2019ehk with complete $\gamma$-ray trapping. The colorbar gradient demonstrates the decline rate dependency on trapping timescale, $t_{\gamma}$ in a radioactive decay powered model fit to the SN~2019ehk data. (b) Comparison between SN~2019ehk and H-poor SN varieties such as SNe~IIb (SN~1993J; light blue squares, \citealt{zhang04}), SNe~Ic-BL (SN~2002ap; light blue plus signs, \citealt{tomita06}), SNe~Ic (SN~2007gr; light blue polygons, \citealt{valenti08a}) and SNe~Ib (iPTF13bvn; light blue stars, \citealt{sri14, fremling16}). The colorbar gradient is based on a fit to SN~2019ehk's light curve and thus cannot be used to directly compare $t_{\gamma}$ values for different SNe.  \label{fig:LC_comparison} }
\end{figure*}

\section{Discussion} \label{sec:discussion}
\subsection{Comparison to Late-time SN Studies} \label{subsec:comparison}

SN~2019ehk is the only confirmed \ca \ to be observed long enough after explosion so as to probe the effects of energy deposition from multiple radioactive isotopes on the bolometric light curve. Studying this object at such late-times also allows for the first test of $\gamma$-ray and positron trapping in a SN that exhibits the typical spectroscopic and photometric evolution of a \ca. As shown in Figure \ref{fig:LC_comparison}(a), the peculiar, ``Calcium-strong'' SN~2016hnk was observed to $\sim$300~days after explosion and \cite{wjg19} found that including $t_{\gamma} \approx 60$~days was necessary to fit the bolometric light curve at late-times. However, while SN~2016hnk follows a similar light curve evolution as SN~2019ehk out to late-times, it is not considered a typical \ca \ given its similarities to ``1991bg-like'' SNe Ia \citep{galbany19}. 

With regards to other thermonuclear SN varieties shown in Figure \ref{fig:LC_comparison}(a), SN~2019ehk has a similar overall light curve evolution out to $\sim$400~days after explosion. Compared to normal and sub-luminous SN Ia, 2011fe and 2005ke, respectively, SN~2019ehk has a consistent decline rate. Furthermore, given the differences in SN parameters e.g., $M_{ej}$ and $E_k$ between SN types, it is understandable that the estimated trapping timescale ($t_{\gamma} \propto M_{ej}/E_k^{1/2}$; Eqn. \ref{eq:F}) is higher in SN~2019ehk ($\sim$54~days) than in SNe Ia ($\lesssim 35$~days). The difference in $\gamma$-ray trapping between objects is illustrated through the $t_{\gamma}$ colorbar in Figure \ref{fig:LC_comparison}. Finally, SN~2019ehk shows a slightly faster bolometric decline than ``Calcium-strong'' SN~2016hnk and Type Iax SN (SN Iax) 2005hk \citep{sahu08}, which suggests that it has less efficient $\gamma$-ray trapping than these low luminosity thermonuclear events. For a phase range of $100-300$~days, SN~2019ehk has a smaller fractional change in luminosity than SN~2005ke but a larger change than all other thermonuclear transients used for comparison. 

We explore the similarities between SN~2019ehk and SNe Iax given that it is a low luminosity transient where the explosion of a WD is a favored progenitor scenario (WJG20a). At late times, we find no evidence for significant change in color to IR bands nor a late-time flattening of the light curve that deviates from a Ni-powered decline, as seen in some SNe~Iax (e.g., SN~2005hk and 2014dt). Furthermore, unlike SNe~Iax, SN~2019ehk has no detectable excess of NIR/MIR flux, which has been used as evidence for both a super-Eddington wind launched from a surviving, bound remnant star (e.g., \citealt{foley16,shen17, Kawabata18}) or dust formation \citep{fox16} in SN Iax, 2014dt. Such scenarios are disfavored based on SN~2019ehk's late-time light curve evolution. 

SN~2019ehk has a faster decline rate and larger fractional change in luminosity ($100-300$~days) than all stripped-envelope SN varieties such as Type IIb, Ibc and broad-lined Ic SNe. As illustrated by Figure \ref{fig:LC_comparison}(b), all of the example H-poor SN sub-types are more efficient at trapping $\gamma$-rays in their ejecta than SN~2019ehk. This suggests values of $t_{\gamma}$ that are a factor of two greater than that of SN~2019ehk and is consistent with the trapping timescales $ \gtrsim 100$~days found by \cite{wheeler15} for a sample of stripped-envelope events. Furthermore, all SNe in that study have larger observed $M_{ej}$ and $E_k$ values than SN~2019ehk despite some objects having comparable total masses of \niVI. 

\subsection{CSM Interaction and Dust Formation} \label{subsec:csm_dust}

SN~2019ehk represents the first \ca \ with direct evidence for confined CSM surrounding the progenitor star at the time of explosion (WJG20a). Shock-ionized spectral lines and luminous X-ray emission revealed that the CSM was H- and He-rich, had a mass of $\sim$ $7\times10^{-3}$~$\Msun$ and a velocity of $\sim$~500~$\kms$. These observations jointly confirmed that this compact shell of material extended out to a radius of $r < 10^{15}$~cm from the progenitor and had a density of $n\approx10^{9}\,\rm{cm^{-3}}$ ($\dot M<10^{-5}\,\rm{M_{\odot}yr^{-1}}$). Radio observations from $\sim$~$30-220$~days after explosion indicated a significantly lower density $n < 10^{4}\,\rm{cm^{-3}}$ at larger radii $r>(0.1-1)\times10^{17}\,\rm{cm}$. Furthermore, there is no evidence of circumstellar interaction in the latest nebular spectrum of SN~2019ehk at a phase of $\sim$270~days. 

Prior to the late-time imaging presented in this analysis, all multi-wavelength observations have indicated that the material ejected by the progenitor was dense, small in quantity, and confined to the immediate circumstellar environment. Based on the light curve modeling in Section \ref{subsec:modeling}, we find no statistical evidence for a power source in addition to \niVI; the same amount of \niVI \ that powers the early-time light curve is enough to account for the entire bolometric luminosity up to 400 days. Here, we quantify the contribution of CSM interaction to the late-time light curve, employing a modified version of the simplified formalism by \cite{smith10}:

\begin{equation} \label{eq:Lcsm}
    L_{\rm CSM} = \frac{1}{2} \epsilon w v_s^3
\end{equation}

where $\epsilon$ is the efficiency of conversion of shock kinetic energy into radiation and $v_s$ is the shock velocity. The wind density parameter $w$ is defined by $\Dot{M}/v_w$, where we adopt $v_w \approx 500$~$\kms$ as estimated in WJG20a. For the explosion parameters of SN~2019ehk ($M_{ej}$ and $E_k$), and a wind-like environment (see WJG20a), the shock velocity is: 

\begin{equation} \label{eq:vs}
    v_s = 10^4 \Bigg [ \Big(\frac{\Dot{M}}{\Msun \ \rm{yr}^{-1} } \Big)^{0.12} \Big(\frac{t}{\rm days} \Big)^{0.12} \Bigg ]^{-1} \ \kms
\end{equation}

We treat Eqn. \ref{eq:Lcsm} as an extra energy term to be added to Eqn. \ref{eq:Lneb} and we derive a 3$\sigma$ limit on mass-loss of $\Dot{M} \lesssim 10^{-10}$~$\Msun \rm yr^{-1}$ for an optimistic efficiency of 80\%. This result indicates very low densities in the SN environment of $n < 10^{4}\,\rm{cm^{-3}}$, consistent with the radio non-detections. Our mass-loss estimate suchs a ``very clean'' environment that is natural in a WD+WD system (WJG20a) but more difficult to reconcile with the environments around massive stars. 

Finally, we consider the case of dust formation in SN~2019ehk for completeness. As shown in the optical/IR light curve in Figure \ref{fig:LCs}(a), the late-time SED of SN~2019ehk is gradually being shifted bluewards, which is not reflective of dust formation that would induce the opposite effect and is likely an effect of fading \ion{Ca}{ii} emission at redward wavelengths. Furthermore, our WFC3/IR observations at +389~days after explosion extend from $\sim$~$0.9-1.7$~$\mu$m and would be able to detect emission from a dust shell that typically peaks around $\sim$~2~$\mu$m. Consequently, we can conclude that there is no evidence for dust formation in SN~2019ehk at phases $\lesssim$~400~days after explosion. 

\subsection{Oxygen Ejecta Mass} \label{subsec:Omass}

The mass of oxygen in the ejecta can constrain the type of progenitor and the explosion mechanism. WJG20a estimated $M_O > 0.14~\Msun$ from \ion{O}{i} and [\ion{O}{i}] emission lines in the +72~day spectrum and \cite{de20b} found a less stringent, but consistent mass limit of $M_O > 0.005 - 0.05~\Msun$ from a spectrum at +270~days using only the [\ion{O}{i}] line transition. Both of these analyzes assumed temperatures of the emission region (e.g., T = 5000~K by WJG20a and T = 3400 - 4000~K by \citealt{de20b}) that were not directly constrained by the data. Here we re-analyze the +72~day spectrum, adding to our analysis the inferences made from an estimated upper limit on the [\ion{O}{i}] 5577 luminosity to constrain the temperature and obtain a robust lower limit to $M_O$.  We then present two independent estimates of $M_O$ based on the $\rm ^{56}Co$ mass obtained in this paper.

\subsubsection{Lower limit from +72~day spectrum}

In order to obtain a lower limit to $M_O$ from the +72~day spectrum, we use $L_{6300}$, the [\ion{O}{i}] $\lambda \lambda$6303,6363 luminosity and $L_{7774}$, the recombination line luminosity from WJG20a. We add a constraint for the [\ion{O}{i}] $\lambda$5577 line by re-scaling to the +44~day spectrum in order to determine the continuum, resulting in a line ratio of $L_{5577}/L_{6300}~<~0.2$ and assuming a constant continuum shape between epochs. We note that a change in the line ratio of $\pm 30 \%$ would influence the excitation rate in $\lambda$6303 by $\pm 25\%$, which in turn will modify the final O mass estimate by $\mp 25\%$. We then computed a grid of models over a range of density, temperature and ionization fraction for various values of the oxygen mass using atomic rates from CHIANTI \citep{dere97}. We find that an O mass of $M_O~>~0.08~\Msun$ is required to match the observed line luminosities, which lies in the upper end of the lower limits presented by \cite{de20b}. For this minimum mass, the other parameters are constrained to log($n_e$) = 8.6, $T = 5350$~K and $\rm O^+/O \sim 0.25$.  For larger $M_O$, wider ranges of the other parameters are allowed. This is a robust lower limit on the O mass because (i) some oxygen could be inside the photosphere at this stage as the spectrum is not yet fully nebular, and (ii) we assume a single temperature. If, as is likely, a range of temperatures is present and the higher temperature gas is more highly ionized, then both the neutral mass ($M \propto e^{22800/T}$) and the ionized mass ($M~\propto~T^{1/2})$ will increase. We note that we can not confirm the approximated formula by \cite{uomoto86} used by \cite{de20b} with CHIANTI. Using the same parameters as \cite{de20b} in CHIANTI, we would infer an O mass that is a factor 1.6 lower than that reported by \cite{de20b}. We speculate that updated atomic parameters of CHIANTI might be responsible for the difference.

\subsubsection{Estimate from $L_{7774}$ in +270~day spectrum}

We measure a \ion{O}{i} recombination line luminosity $L_{7774} = 1.8\times 10^{37}~\rm erg~s^{-1}$ in the +270 spectrum. It is known that  $\sim$~37\% of the recombinations produce that line, and each recombination must balance an ionization \citep{julienne74}. We have shown that the original mass of $\rm ^{56}Co$ is 0.028 $\Msun$, but by +270~days, only 0.0025~$\Msun$ remains.  With a 77.2~day half-life, that implies $5.4 \times 10^{45}$ decays per second at $t = 270$~days, each of which carries 2.11 MeV of energy. \cite{victor94} computed the number of ionizations per 1000~eV as a particle slows down in pure oxygen gas. They did not include photo-ionization by emission lines created in the process and while the \ion{O}{i} emission lines cannot photo-ionize oxygen, \ion{O}{ii} lines such as those at 834~\AA\/ can ionize \ion{O}{i}. We use this information to quantify the amount of energy released by \coVI \ decay that is channeled solely to O I emission. We note that Ca is excited by a population of electrons at significantly lower energies that would not lead to O emission. Adding in those photo-ionizations, we find $26-45$ ionizations per 1000~eV.  For a radius of $6 \times 10^{15}$~cm based on the expansion speed and phase, the energy flux is $2.7 \times 10^{13}$ MeV$\rm ~cm^{-2}~s^{-1}$, and the absorption cross section based on $\kappa_{\gamma}$ = 0.027 $\rm cm^2~g^{-1}$ yields $(5.1-8.7)\times 10^{-7}$ ionizations per second.  Thus the observed $\lambda$7774 luminosity requires $M_O \approx 0.30-0.50$~$\Msun$.  This estimate applies if the $^{56}$Co is located well inside the absorbing shell but the local $\gamma$-ray flux will be higher if the $^{56}$Co is just inside the absorbing shell (i.e., large degree of mixing). The geometrical correction could reduce the required oxygen mass by as much as a factor of 1.5, and the final estimate is $M_O \approx 0.20~-~0.33~\Msun$. We also explore the effect of a large host extinction on the SN~2019ehk O mass. Using a color excess of $E(B-V) = 1.0$~mag, we find a \ion{O}{i} line luminosity $L_{7774} = 4.3\times 10^{37}~\rm erg~s^{-1}$ which yields an O mass range of $M_O \approx 0.70-1.20$~$\Msun$ by the steps outlined above. While these values violate the total ejecta mass estimates from light curve modeling and support a lower host extinction value, it is possible that the assumptions made in this analysis do not fully account for all the details of SN physics e.g., the application of \cite{victor94} is technically for pure O gas.


\subsubsection{Estimate from opacity at +270~days}

Figure \ref{fig:LCs}(b) shows that all but $\sim$4\% of the radioactive decay energy escapes from the oxygen SN ejecta shell, which indicates an optical depth of $\sim$0.04. We assume that the source of $\gamma$-rays (i.e., \coVI) is centrally located. With an opacity $\kappa_{\gamma} = 0.027 ~\rm cm^2~g^{-1}$, that implies a mass column of 1.48~g~$\rm cm^{-2}$.  Multiplying by the area of a $6 \times 10^{15}$~cm shell gives an O mass of $\sim$~0.3~$\Msun$. WJG20a found that there is a significant amount of He in the ejecta, which would reduce the O mass range to $\sim$~$0.27~\Msun$. Carbon might be present as well, which could lower the O mass by as much as 1/3. 

A further geometric correction should also be considered.  The estimate above implicitly assumes that the $\gamma$-rays move radially, and that is a good approximation if the $^{56}$Co is located well inside the absorbing shell. If the $^{56}$Co is located just inside the absorbing shell, a photon will move at some angle to the radial and will encounter more material.  The correction factor depends on the thickness of the shell, but for a plausible range of $1.5 ~< ~r_{outer}/r_{inner} ~<~ 2$, the mass estimate could be decreased by a factor of 1.5 to 1.32.  If the $^{56}$Co is mixed with the absorbing material, some $\gamma$-rays will escape more easily, bringing the correction factor back toward 1. The mass estimate based on the opacity thus becomes $\sim$~$0.20~\Msun$.
This O mass, as well as other estimates discussed above, are consistent with the O abundances in merging hybrid + C/O WDs (e.g., \citealt{Zenati2019}) that WJG20a present as a favored progenitor scenario for SN~2019ehk.  

\subsection{Progenitor Channels} \label{subsec:progenitors}

As the sample of known \cas \ continues to grow, the exact progenitor systems responsible for these SNe remains unknown. While the older stellar environments and significant host galaxy offsets observed for many \cas \ is consistent with a WD origin, the increasing diversity of \ca \ explosion sites indicates that their progenitors may be heterogeneous and include some types of massive stars. For the progenitor of SN~2019ehk, \cite{Nakaoka20} conclude that the SN may have arisen from the explosion of a ultra-stripped, low mass He (or C/O) star in a binary system with a companion NS. Alternatively, WJG20a find that the disruption of a hybrid WD by a C/O WD companion is most consistent with observations. However, as identified in WJG20a, pre-explosion imaging also allows for a low-mass, massive star progenitor ($\sim$10~$\Msun$) in a binary system.  

\begin{figure}[t!]
\centering
\includegraphics[width=0.45\textwidth]{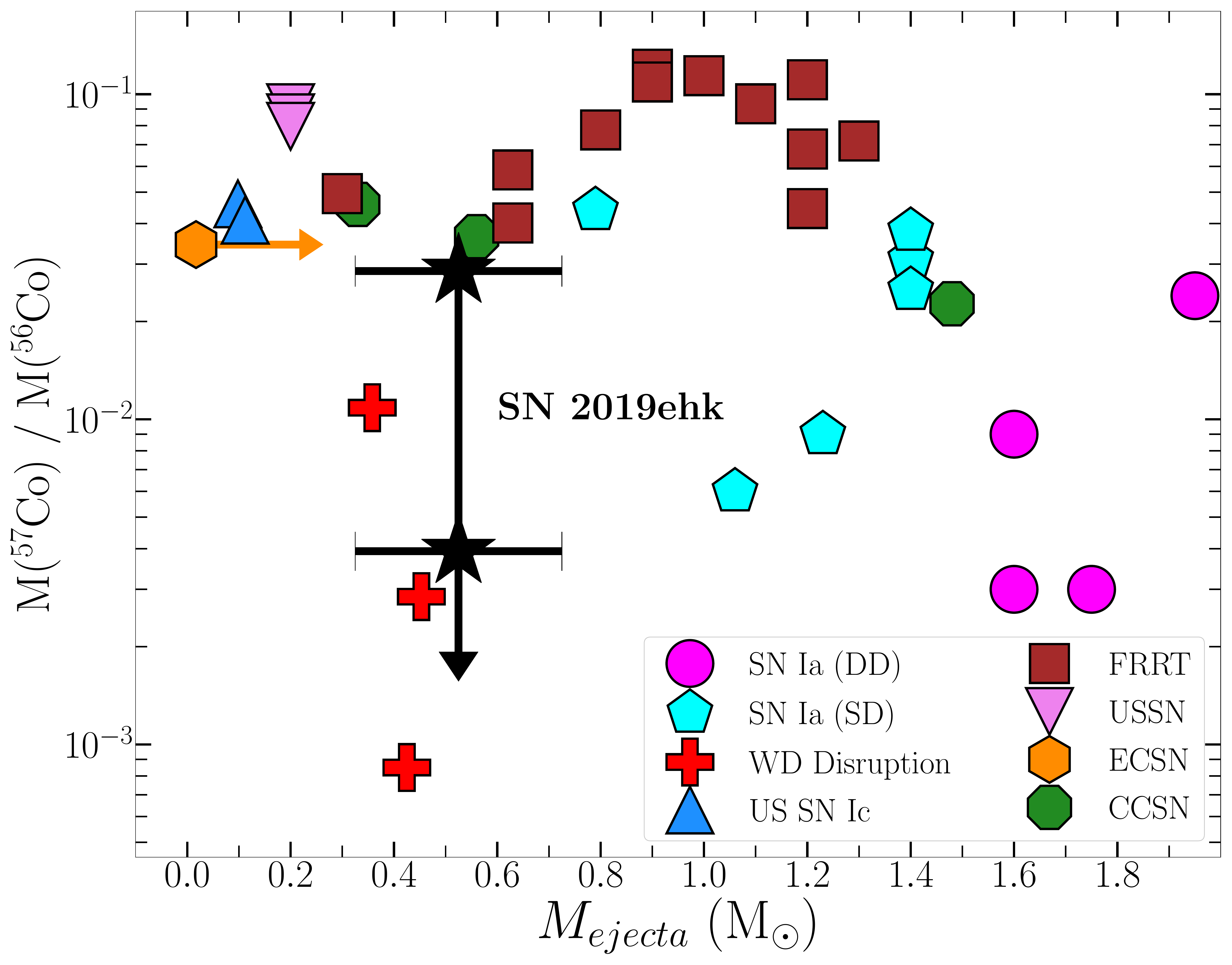}
\caption{Comparison of cobalt isotope mass ratio to $M_{ej}$ for SN~2019ehk (black star) with respect to values predicted by various explosion models. SN~2019ehk's mass ratio is shown as range of most to least conservative limits based on no trapping to complete $\gamma$-ray trapping, respectively, in \coVII \ decay. These limits are not dependent on the assumed host galaxy extinction (e.g., see \S\ref{subsec:modeling}). Single and double degenerate SN Ia-like explosions are presented as cyan polygons and magenta circles, respectively. WD disruption models for \cas \ are shown as red plus signs and different ultra-stripped (US) SN models presented as blue/pink up/down triangles. Model for WD+NS/BH disruptions leading to a Faint Rapid Red Transient (FRRT) are shown as brown squares. Electron-capture (EC) and core-collapse (CC) SN models for low mass progenitors ($\lesssim$~11~$\Msun$) are shown as orange hexagons and green octagons, respectively. All model parameters presented in Table \ref{tab:explosion_models}. The uncertainty on $M_{ej}$ for SN~2019ehk reflects the range of values estimated in both WJG20a and \cite{Nakaoka20}. \label{fig:ratio}}
\end{figure}

In the context of late-time studies of SNe Ia, the ratio of odd to even radioactive isotope masses (e.g., ${}^{57}\textrm{Co}$ / ${}^{56}\textrm{Co}$) provides information on the density structure of the explosion which, in turn, can help constrain the progenitor system of these SNe \citep{seitenzahl13a, Graur16, shappee17, dimitriadis17, jacobson-galan18}. Here, we use the mass ratio $M({}^{57}\textrm{Co}) / M({}^{56}\textrm{Co})$ derived in \S\ref{subsec:modeling} as a unique and novel probe of possible \ca \ progenitor scenarios. In Figure \ref{fig:ratio}, we compare SN~2019ehk's radioactive isotope mass ratio limit and total ejecta mass to those predicted in a variety of explosion models. The complete list of models used in this plot are presented in Table \ref{tab:explosion_models}. It should be noted that the complexity of the nuclear reaction network may vary between each type of explosion model.  

With respect to single degenerate and double degenerate models for SN Ia-like explosions of WDs, SN~2019ehk has a lower total ejecta mass but is consistent with the mass ratio of some explosion models. A progenitor scenario with similar nucleosynthesis that instead involved the explosion of a sub-Chandrasekhar mass WD (e.g., He-shell double detonations) would match these specific observables in SN~2019ehk. However, any WD explosion model of this variety would need to reconcile the H- and He-rich CSM in near SN~2019ehk's progenitor star (WJG20a). 

More exotic progenitor scenarios such as disruptions of low-mass WDs by another WD or a NS can also be constrained. As shown in Figure \ref{fig:ratio}, SN~2019ehk is inconsistent with the predicted ejecta masses and nucleosynthesis of CO or ONe WD + NS merger models (\citealt{zenati20}, Bobrick et al. 2020b, in prep.). However, the exact unbound ejecta mass produced in these models is uncertain and could match $M_{ej}$ estimated for SN~2019ehk. Nonetheless, most of these models synthesize higher amounts of \coVII \ than could be present in SN~2019ehk (Fig. \ref{fig:ratio}) and thus are not viable progenitor systems.

An explosion scenario that is consistent with SN~2019ehk is the disruption of a lower-mass CO WD (or hybrid HeCO WD) by a hybrid WD (\citealt{Zenati2019}), which can produce fast-rising, faint \ca-like events (Zenati et al. 2020b, in prep.). The explosion model can result in $\sim$0.4 - 0.6~$\Msun$ of ejecta and synthesize low enough masses of \coVII \ so as to remain within the limit set by the late-time light curve modeling. WJG20a also find this progenitor scenario to be consistent with the H- and He-rich CSM composition found in the SN progenitor environment. While this late-time analysis confirms one of the favored models in WJG20a, further tests should be done to understand how well this type of explosion can quantitatively match SN~2019ehk's early-time light curve and spectra. 

Lastly, we compare estimates of $M_{ej}$ and mass ratio values to those predicted by a variety of core-collapse (CC) SN models. In Figure \ref{fig:ratio}, we show that SN~2019ehk is not consistent with both the nucleosynthetic yields and ejecta masses produced in the explosion of massive stars in the lowest mass bins ($\lesssim$~$8-11$~$\Msun$; \citealt{wanajo18}). Similarly, electron-capture (EC) SN models for low mass progenitors (8.8~$\Msun$; \citealt{wanajo18}) cannot reproduce the SN~2019ehk observables despite their proposed link to fastly evolving transients such as \cas \ \citep{moriya16, milisavljevic17}. Furthermore, we explore the possible connection between SN~2019ehk and ultra-stripped SN (USSN) models. Such a progenitor system was favored by \cite{Nakaoka20} for SN~2019ehk and involves the explosion of a He or CO star that has had most of its outer H and He envelope removed due to its NS companion. In Figure \ref{fig:ratio} we include explosion models for ultra-stripped SNe Ic \citep{yoshida17} and USSNe of varying explosion energies \citep{moriya17}. Both models produce less $M_{ej}$ than SN~2019ehk and synthesize too much \coVII \ to be consistent with the most conservative mass ratio limit. Nonetheless, additional modeling of USSNe is needed to understand the range of ejecta and isotope masses generated through the explosion of compact, stripped massive stars.

Based on a lower limit on the O mass of $M_O > 0.005 - 0.05$~$\Msun$, \cite{de20b} favor a stripped envelope progenitor ($M_{\rm ZAMS}\approx$~$9-9.5$~$\Msun$) for SN~2019ehk  with $\gtrsim 0.02$~$\Msun$ of H on the stellar surface. The mass of potential photospheric H is based on a qualitative analogy between one peak spectrum of SN~2019ehk and SNe~IIb models (e.g., \citealt{hachinger12}) despite the overall dissimilarity between the photometric and spectroscopic evolution, as well as explosion parameters, of H-poor SNe compared to SN~2019ehk (\citealt{Nakaoka20}, WJG20a). While the O mass lower limit by \cite{de20b} shows consistency with USSN models (e.g., \citealt{moriya17, yoshida17}), such a progenitor scenario is inconsistent with even the most conservative mass ratio limit shown in Figure \ref{fig:ratio} as well as alternative methods for calculating O mass discussed in Section \ref{subsec:Omass}. Furthermore, the range of nucleosynthetic yields and $M_{ej}$ values produced in the core-collapse of normal to ultra-stripped, $9-11$~$\Msun$ SN progenitors cannot reproduce those observed in SN~2019ehk.

Additionally, it is difficult to reconcile the specific progenitor scenario proposed by \cite{de20b} with the detection of a dense, confined shell of H- and He-rich CSM in the SN~2019ehk progenitor environment. From X-ray detections and flash-ionized spectral lines, WJG20a derived a CSM H mass of $\sim$~$3 \times 10^{-4} \ \Msun$ around SN~2019ehk's progenitor, which is incompatible with the estimate proposed by \cite{de20b} of $\gtrsim$~0.03~$\Msun$ near or on the surface of the progenitor. Furthermore, the progenitor environment of SN~2019ehk is unlike that of any double-peaked, H-poor SNe with extensive radio observations (e.g., see \citealt{kamble16}) and the lack of radio detections indicates a low density environment at distances $r > 10^{16}$~cm, which is inconsistent with a stripped-envelope, massive star progenitor system. Also, our radio limits, as well as the presence of a H-rich CSM, are inconsistent with most of the ultra-stripped SN progenitor configurations presented by \cite{Matsuoka20}. Nonetheless, SN~2019ehk radio limits and ejecta mass are consistent with two binary models that include a fraction of gas escaping the system $f_{\dot{M}} = 0.1$ and final orbital separation $a_{\rm fin} = 1-10~\Rsun$, but these models cannot reconcile the presence of H in the local SN environment. Furthermore, we note the large uncertainties on the efficiency of non-conservative mass transfer in these systems during the Roche lobe overflow stage of binary evolution. It is unclear how mass-loss in a stripped, $\sim$~$9-9.5$~$\Msun$ massive star progenitor could allow for the presence of  only $\sim 10^{-4}$~$\Msun$ of dense H-rich material in its local environment ($r < 10^{15}$~cm), while also ejecting several $\Msun$ worth of H via binary interaction that was not detected in any panchromatic observations of SN~2019ehk out to late-time phases. 

\section{Conclusions} \label{sec:conclusion}

In this paper, we have presented  \textit{HST} WFC3 imaging of \ca \ SN~2019ehk at 276-389~days after explosion. Photometric detections in all optical/IR filters enabled the creation of a bolometric light curve that extended out to phases yet unexplored in a \ca. We show that the late-time light curve evolution can be modeled solely through the radioactive decay of isotope \coVI \ with a mass of $M({}^{56}\textrm{Co}) = (2.8 \pm 0.1) \times 10^{-2}$~$\Msun$. Additionally, we find evidence for $\gamma$-ray leakage on the timescale of $t_{\gamma} = 53.9 \pm 1.3$ d but do not find statistical evidence for incomplete positron trapping in SN~2019ehk's ejecta. The bolometric light curve of SN~2019ehk does not extend to late enough phases to precisely quantify the mass of \coVII \ synthesized in the explosion and therefore we derive the most conservative limit on the mass ratio of odd to even isotopes to be $M({}^{57}\textrm{Co}) / M({}^{56}\textrm{Co}) \leq 0.030$.

We compare this mass ratio limit and the total SN~2019ehk ejecta mass to that predicted by various explosions models involving WDs and stripped, compact massive stars. We show that these observables make SN~2019ehk incompatible with single- and double-degenerate explosion scenarios that typically produce SN Ia-like explosions. Additionally, SN~2019ehk is inconsistent with the projected nucleosynthetic yields of WD+NS binary mergers as well as CC and EC SNe from normal to ultra-stripped massive stars ($M_{\rm{ZAMS}}\approx$~9-11~$\Msun$). However, these derived values in SN~2019ehk do match the mass ratio and $M_{ej}$ produced in the tidal disruption of a low-mass C/O WD by a larger, hydrid WD. Additional modeling of these explosion mechanisms, as well as more late-time observations of nearby \cas, will be essential in constraining \ca \ progenitor systems. 

\section{Acknowledgements} \label{Sec:ack}

Research at Northwestern University and CIERA is conducted on the stolen land of the Council of Three Fires, the Ojibwe, Potawatomi, and Odawa people, as well as the Menominee, Miami and Ho-Chunk nations. 

We thank Takashi Yoshida, Takashi Moriya and Shinya Wanajo for model yields. W.J-G is supported by the National Science Foundation Graduate Research Fellowship Program under Grant No.~DGE-1842165 and the IDEAS Fellowship Program at Northwestern University. W.J-G acknowledges support through NASA grants in support of {\it Hubble Space Telescope} program GO-16075. R.M. is grateful to KITP for hospitality during the completion of this paper. This research was supported in part by the National Science Foundation under Grant No. NSF PHY-1748958. R.M. acknowledges support by the National Science Foundation under Award No. AST-1909796. Raffaella Margutti is a CIFAR Azrieli Global Scholar in the Gravity \& the Extreme Universe Program, 2019. The Margutti's team at Northwestern is partially funded by the Heising-Simons Foundation under grant \# 2018-0911 (PI: Margutti). C.D.K. acknowledges support through NASA grants in support of {\it Hubble Space Telescope} programmes GO-15691 and AR-16136. P.K.B. is supported by a CIERA Postdoctoral Fellowship. HBP acknowledges support from the Kingsely distinguished-visitor program at Caltech,  the KITP visitor program (supported in part by the National Science Foundation under Grant No. NSF PHY-1748958), and the European Union's Horizon 2020 research and innovation program under grant agreement No 865932-ERC-SNeX.

The UCSC transient team is supported in part by NSF grant AST-1518052, NASA/{\it Swift} grant 80NSSC19K1386, the Gordon \& Betty Moore Foundation, the Heising-Simons Foundation, and by a fellowship from the David and Lucile Packard Foundation to R.J.F. Research at Lick Observatory is partially supported by a generous gift from Google.

This paper includes data gathered with the 6.5 meter Magellan Telescopes located at Las Campanas Observatory, Chile. CHIANTI is a collaborative project involving George Mason University, the University of Michigan (USA), University of Cambridge (UK) and NASA Goddard Space Flight Center (USA).

\facilities{\emph{Hubble Space Telescope}}

\software{scipy \citep{virtanen20}, IRAF (Tody 1986, Tody 1993), AstroDrizzle \citep{astrodrizzle}, photpipe \citep{Rest+05}, DoPhot \citep{Schechter+93}, HOTPANTS \citep{becker15}}, dolphot \citep{dolphot}

\bibliographystyle{aasjournal} 
\bibliography{references}

\begin{thebibliography}{}
\expandafter\ifx\csname natexlab\endcsname\relax\def\natexlab#1{#1}\fi
\providecommand{\url}[1]{\href{#1}{#1}}
\providecommand{\dodoi}[1]{doi:~\href{http://doi.org/#1}{\nolinkurl{#1}}}
\providecommand{\doeprint}[1]{\href{http://ascl.net/#1}{\nolinkurl{http://ascl.net/#1}}}
\providecommand{\doarXiv}[1]{\href{https://arxiv.org/abs/#1}{\nolinkurl{https://arxiv.org/abs/#1}}}

\bibitem[{{Arnett}(1982)}]{arnett82}
{Arnett}, W.~D. 1982, \apj, 253, 785, \dodoi{10.1086/159681}

\bibitem[{{Becker}(2015)}]{becker15}
{Becker}, A. 2015, {HOTPANTS: High Order Transform of PSF ANd Template
  Subtraction}, Astrophysics Source Code Library.
\newblock \doeprint{1504.004}

\bibitem[{{Bobrick} {et~al.}(2017){Bobrick}, {Davies}, \&
  {Church}}]{Bobrick2017}
{Bobrick}, A., {Davies}, M.~B., \& {Church}, R.~P. 2017, \mnras, 467, 3556,
  \dodoi{10.1093/mnras/stx312}

\bibitem[{{Cappellaro} {et~al.}(1997){Cappellaro}, {Mazzali}, {Benetti},
  {Danziger}, {Turatto}, {della Valle}, \& {Patat}}]{Cappellaro97}
{Cappellaro}, E., {Mazzali}, P.~A., {Benetti}, S., {et~al.} 1997, \aap, 328,
  203.
\newblock \doarXiv{astro-ph/9707016}

\bibitem[{{Clocchiatti} \& {Wheeler}(1997)}]{clocchiatti97}
{Clocchiatti}, A., \& {Wheeler}, J.~C. 1997, \apj, 491, 375,
  \dodoi{10.1086/304961}

\bibitem[{{Colgate} {et~al.}(1980){Colgate}, {Petschek}, \&
  {Kriese}}]{colgate80}
{Colgate}, S.~A., {Petschek}, A.~G., \& {Kriese}, J.~T. 1980, \apjl, 237, L81,
  \dodoi{10.1086/183239}

\bibitem[{{Contreras} {et~al.}(2010){Contreras}, {Hamuy}, {Phillips},
  {Folatelli}, {Suntzeff}, {Persson}, {Stritzinger}, {Boldt}, {Gonz{\'a}lez},
  {Krzeminski}, {Morrell}, {Roth}, {Salgado}, {Maureira}, {Burns}, {Freedman},
  {Madore}, {Murphy}, {Wyatt}, {Li}, \& {Filippenko}}]{contreras10}
{Contreras}, C., {Hamuy}, M., {Phillips}, M.~M., {et~al.} 2010, \aj, 139, 519,
  \dodoi{10.1088/0004-6256/139/2/519}

\bibitem[{{De} {et~al.}(2020{\natexlab{a}}){De}, {Fremling}, {Gal-Yam},
  {Kasliwal}, \& {Kulkarni}}]{de20b}
{De}, K., {Fremling}, U.~C., {Gal-Yam}, A., {Kasliwal}, M.~M., \& {Kulkarni},
  S.~R. 2020{\natexlab{a}}, arXiv e-prints, arXiv:2009.02347.
\newblock \doarXiv{2009.02347}

\bibitem[{{De} {et~al.}(2020{\natexlab{b}}){De}, {Kasliwal}, {Tzanidakis},
  {Fremling}, {Adams}, {Andreoni}, {Bagdasaryan}, {Bellm}, {Bildsten},
  {Cannella}, {Cook}, {Delacroix}, {Drake}, {Duev}, {Dugas}, {Frederick},
  {Gal-Yam}, {Goldstein}, {Golkhou}, {Graham}, {Hale}, {Hankins}, {Helou},
  {Ho}, {Irani}, {Jencson}, {Kaye}, {Kulkarni}, {Kupfer}, {Laher},
  {Leadbeater}, {Lunnan}, {Masci}, {Miller}, {Neill}, {Ofek}, {Perley},
  {Polin}, {Prince}, {Quataert}, {Reiley}, {Riddle}, {Rusholme}, {Sharma},
  {Shupe}, {Sollerman}, {Tartaglia}, {Walters}, {Yan}, \& {Yao}}]{de20}
{De}, K., {Kasliwal}, M.~M., {Tzanidakis}, A., {et~al.} 2020{\natexlab{b}},
  arXiv e-prints, arXiv:2004.09029.
\newblock \doarXiv{2004.09029}

\bibitem[{{Dere} {et~al.}(1997){Dere}, {Landi}, {Mason}, {Monsignori Fossi}, \&
  {Young}}]{dere97}
{Dere}, K.~P., {Landi}, E., {Mason}, H.~E., {Monsignori Fossi}, B.~C., \&
  {Young}, P.~R. 1997, \aaps, 125, 149, \dodoi{10.1051/aas:1997368}

\bibitem[{{Dimitriadis} {et~al.}(2017){Dimitriadis}, {Sullivan}, {Kerzendorf},
  {Ruiter}, {Seitenzahl}, {Taubenberger}, {Doran}, {Gal-Yam}, {Laher},
  {Maguire}, {Nugent}, {Ofek}, \& {Surace}}]{dimitriadis17}
{Dimitriadis}, G., {Sullivan}, M., {Kerzendorf}, W., {et~al.} 2017, \mnras,
  468, 3798, \dodoi{10.1093/mnras/stx683}

\bibitem[{{Dolphin}(2000)}]{dolphot}
{Dolphin}, A.~E. 2000, \pasp, 112, 1383, \dodoi{10.1086/316630}

\bibitem[{{Dressler} {et~al.}(2011){Dressler}, {Bigelow}, {Hare}, {Sutin},
  {Thompson}, {Burley}, {Epps}, {Oemler}, {Bagish}, \& {Birk}}]{dressler11}
{Dressler}, A., {Bigelow}, B., {Hare}, T., {et~al.} 2011, \pasp, 123, 288,
  \dodoi{10.1086/658908}

\bibitem[{{Filippenko} {et~al.}(2003){Filippenko}, {Chornock}, {Swift},
  {Modjaz}, {Simcoe}, \& {Rauch}}]{filippenko03}
{Filippenko}, A.~V., {Chornock}, R., {Swift}, B., {et~al.} 2003, \iaucirc,
  8159, 2

\bibitem[{{Fink} {et~al.}(2014){Fink}, {Kromer}, {Seitenzahl},
  {Ciaraldi-Schoolmann}, {R{\"o}pke}, {Sim}, {Pakmor}, {Ruiter}, \&
  {Hillebrandt}}]{Fink14MNRAS}
{Fink}, M., {Kromer}, M., {Seitenzahl}, I.~R., {et~al.} 2014, \mnras, 438,
  1762, \dodoi{10.1093/mnras/stt2315}

\bibitem[{{Fitzpatrick}(1999)}]{fitzpatrick99}
{Fitzpatrick}, E.~L. 1999, \pasp, 111, 63, \dodoi{10.1086/316293}

\bibitem[{{Foley}(2015)}]{foley15}
{Foley}, R.~J. 2015, \mnras, 452, 2463, \dodoi{10.1093/mnras/stv789}

\bibitem[{{Foley} {et~al.}(2016){Foley}, {Jha}, {Pan}, {Zheng}, {Bildsten},
  {Filippenko}, \& {Kasen}}]{foley16}
{Foley}, R.~J., {Jha}, S.~W., {Pan}, Y.-C., {et~al.} 2016, \mnras, 461, 433,
  \dodoi{10.1093/mnras/stw1320}

\bibitem[{{Fox} {et~al.}(2016){Fox}, {Johansson}, {Kasliwal}, {Andrews},
  {Bally}, {Bond}, {Boyer}, {Gehrz}, {Helou}, {Hsiao}, {Masci},
  {Parthasarathy}, {Smith}, {Tinyanont}, \& {Van Dyk}}]{fox16}
{Fox}, O.~D., {Johansson}, J., {Kasliwal}, M., {et~al.} 2016, \apjl, 816, L13,
  \dodoi{10.3847/2041-8205/816/1/L13}

\bibitem[{{Fransson} \& {Jerkstrand}(2015)}]{Fransson15}
{Fransson}, C., \& {Jerkstrand}, A. 2015, \apjl, 814, L2,
  \dodoi{10.1088/2041-8205/814/1/L2}

\bibitem[{{Fransson} \& {Kozma}(1993)}]{frannson93}
{Fransson}, C., \& {Kozma}, C. 1993, \apjl, 408, L25, \dodoi{10.1086/186822}

\bibitem[{{Fremling} {et~al.}(2016){Fremling}, {Sollerman}, {Taddia}, {Ergon},
  {Fraser}, {Karamehmetoglu}, {Valenti}, {Jerkstrand }, {Arcavi}, {Bufano},
  {Elias Rosa}, {Filippenko}, {Fox}, {Gal-Yam}, {Howell}, {Kotak}, {Mazzali},
  {Milisavljevic}, {Nugent}, {Nyholm}, {Pian}, \& {Smartt}}]{fremling16}
{Fremling}, C., {Sollerman}, J., {Taddia}, F., {et~al.} 2016, \aap, 593, A68,
  \dodoi{10.1051/0004-6361/201628275}

\bibitem[{{Galbany} {et~al.}(2019){Galbany}, {Ashall}, {Hoeflich},
  {Gonz{\'a}lez-Gait{\'a}n}, {Taubenberger}, {Stritzinger}, {Hsiao}, {Mazzali},
  {Baron}, {Blondin}, {Bose}, {Bulla}, {Burke}, {Burns}, {Cartier}, {Della
  Valle}, {Diamond}, {Guti{\'e}rrez}, {Harmanen}, {Hiramatsu}, {Holoien},
  {Hosseinzadeh}, {Howell}, {Huang}, {Inserra}, {de Jaeger}, {Jha}, {Kangas},
  {Kromer}, {Lyman}, {Maguire}, {Marion}, {Milisavljevic}, {Prentice}, {Razza},
  {Reynolds}, {Sand}, {Shappee}, {Shekhar}, {Smartt}, {Stassun}, {Sullivan},
  {Valenti}, {Villanueva}, {Wang}, {Wheeler}, {Zhai}, \& {Zhang}}]{galbany19}
{Galbany}, L., {Ashall}, C., {Hoeflich}, P., {et~al.} 2019, arXiv e-prints,
  arXiv:1904.10034.
\newblock \doarXiv{1904.10034}

\bibitem[{{Gonzaga}(2012)}]{astrodrizzle}
{Gonzaga}, S. 2012, {The DrizzlePac Handbook}

\bibitem[{{Graur} {et~al.}(2016){Graur}, {Zurek}, {Shara}, {Riess},
  {Seitenzahl}, \& {Rest}}]{Graur16}
{Graur}, O., {Zurek}, D., {Shara}, M.~M., {et~al.} 2016, \apj, 819, 31,
  \dodoi{10.3847/0004-637X/819/1/31}

\bibitem[{{Graur} {et~al.}(2018){Graur}, {Zurek}, {Rest}, {Seitenzahl},
  {Shappee}, {Fisher}, {Guillochon}, {Shara}, \& {Riess}}]{graur17}
{Graur}, O., {Zurek}, D.~R., {Rest}, A., {et~al.} 2018, \apj, 859, 79,
  \dodoi{10.3847/1538-4357/aabe25}

\bibitem[{{Hachinger} {et~al.}(2012){Hachinger}, {Mazzali}, {Taubenberger},
  {Hillebrand t}, {Nomoto}, \& {Sauer}}]{hachinger12}
{Hachinger}, S., {Mazzali}, P.~A., {Taubenberger}, S., {et~al.} 2012, \mnras,
  422, 70, \dodoi{10.1111/j.1365-2966.2012.20464.x}

\bibitem[{{Iwamoto} {et~al.}(1999){Iwamoto}, {Brachwitz}, {Nomoto},
  {Kishimoto}, {Umeda}, {Hix}, \& {Thielemann}}]{iwamoto99}
{Iwamoto}, K., {Brachwitz}, F., {Nomoto}, K., {et~al.} 1999, \apjs, 125, 439,
  \dodoi{10.1086/313278}

\bibitem[{{Jacobson-Gal{\'a}n} {et~al.}(2018){Jacobson-Gal{\'a}n},
  {Dimitriadis}, {Foley}, \& {Kilpatrick}}]{jacobson-galan18}
{Jacobson-Gal{\'a}n}, W.~V., {Dimitriadis}, G., {Foley}, R.~J., \&
  {Kilpatrick}, C.~D. 2018, \apj, 857, 88, \dodoi{10.3847/1538-4357/aab716}

\bibitem[{{Jacobson-Gal{\'a}n}
  {et~al.}(2020{\natexlab{a}}){Jacobson-Gal{\'a}n}, {Margutti}, {Kilpatrick},
  {Hiramatsu}, {Perets}, {Khatami}, {Foley}, {Raymond}, {Yoon}, {Bobrick},
  {Zenati}, {Galbany}, {Andrews}, {Brown}, {Cartier}, {Coppejans},
  {Dimitriadis}, {Dobson}, {Hajela}, {Howell}, {Kuncarayakti}, {Milisavljevic},
  {Rahman}, {Rojas-Bravo}, {Sand}, {Shepherd}, {Smartt}, {Stacey}, {Stroh},
  {Swift}, {Terreran}, {Vinko}, {Wang}, {Anderson}, {Baron}, {Berger},
  {Blanchard}, {Burke}, {Coulter}, {DeMarchi}, {DerKacy}, {Fremling}, {Gomez},
  {Gromadzki}, {Hosseinzadeh}, {Kasen}, {Kriskovics}, {McCully},
  {M{\"u}ller-Bravo}, {Nicholl}, {Ordasi}, {Pellegrino}, {Piro}, {P{\'a}l},
  {Ren}, {Rest}, {Rich}, {Sai}, {S{\'a}rneczky}, {Shen}, {Short}, {Siebert},
  {Stauffer}, {Szak{\'a}ts}, {Zhang}, {Zhang}, \& {Zhang}}]{wjg20}
{Jacobson-Gal{\'a}n}, W.~V., {Margutti}, R., {Kilpatrick}, C.~D., {et~al.}
  2020{\natexlab{a}}, \apj, 898, 166, \dodoi{10.3847/1538-4357/ab9e66}

\bibitem[{{Jacobson-Gal{\'a}n}
  {et~al.}(2020{\natexlab{b}}){Jacobson-Gal{\'a}n}, {Polin}, {Foley},
  {Dimitriadis}, {Kilpatrick}, {Margutti}, {Coulter}, {Jha}, {Jones},
  {Kirshner}, {Pan}, {Piro}, {Rest}, \& {Rojas-Bravo}}]{wjg19}
{Jacobson-Gal{\'a}n}, W.~V., {Polin}, A., {Foley}, R.~J., {et~al.}
  2020{\natexlab{b}}, \apj, 896, 165, \dodoi{10.3847/1538-4357/ab94b8}

\bibitem[{{Julienne} {et~al.}(1974){Julienne}, {Davis}, \& {Oran}}]{julienne74}
{Julienne}, P.~S., {Davis}, J., \& {Oran}, E. 1974, \jgr, 79, 2540,
  \dodoi{10.1029/JA079i016p02540}

\bibitem[{{Kamble} {et~al.}(2016){Kamble}, {Margutti}, {Soderberg},
  {Chakraborti}, {Fransson}, {Chevalier}, {Powell}, {Milisavljevic}, {Parrent},
  \& {Bietenholz}}]{kamble16}
{Kamble}, A., {Margutti}, R., {Soderberg}, A.~M., {et~al.} 2016, \apj, 818,
  111, \dodoi{10.3847/0004-637X/818/2/111}

\bibitem[{{Kasliwal} {et~al.}(2012){Kasliwal}, {Kulkarni}, {Gal-Yam}, {Nugent},
  {Sullivan}, {Bildsten}, {Yaron}, {Perets}, {Arcavi}, {Ben-Ami}, {Bhalerao},
  {Bloom}, {Cenko}, {Filippenko}, {Frail}, {Ganeshalingam}, {Horesh}, {Howell},
  {Law}, {Leonard}, {Li}, {Ofek}, {Polishook}, {Poznanski}, {Quimby},
  {Silverman}, {Sternberg}, \& {Xu}}]{kasliwal12}
{Kasliwal}, M.~M., {Kulkarni}, S.~R., {Gal-Yam}, A., {et~al.} 2012, \apj, 755,
  161, \dodoi{10.1088/0004-637X/755/2/161}

\bibitem[{{Kawabata} {et~al.}(2018){Kawabata}, {Kawabata}, {Maeda}, {Yamanaka},
  {Nakaoka}, {Takaki}, {Fukushima}, {Kojiguchi}, {Masumoto}, {Matsumoto},
  {Akitaya}, {Itoh}, {Kand a}, {Moritani}, {Takata}, {Uemura}, {Ui}, {Yoshida},
  {Hattori}, {Lee}, {Tominaga}, \& {Nomoto}}]{Kawabata18}
{Kawabata}, M., {Kawabata}, K.~S., {Maeda}, K., {et~al.} 2018, \pasj, 70, 111,
  \dodoi{10.1093/pasj/psy116}

\bibitem[{{Kilpatrick} {et~al.}(2018){Kilpatrick}, {Takaro}, {Foley},
  {Leibler}, {Pan}, {Campbell}, {Jacobson-Galan}, {Lewis}, {Lyke}, {Max},
  {Medallon}, \& {Rest}}]{kilpatrick18}
{Kilpatrick}, C.~D., {Takaro}, T., {Foley}, R.~J., {et~al.} 2018, \mnras, 480,
  2072, \dodoi{10.1093/mnras/sty2022}

\bibitem[{{Kromer} {et~al.}(2013){Kromer}, {Pakmor}, {Taubenberger}, {Pignata},
  {Fink}, {R{\"o}pke}, {Seitenzahl}, {Sim}, \& {Hillebrandt}}]{Kromer13ApJ}
{Kromer}, M., {Pakmor}, R., {Taubenberger}, S., {et~al.} 2013, \apjl, 778, L18,
  \dodoi{10.1088/2041-8205/778/1/L18}

\bibitem[{{Kromer} {et~al.}(2016){Kromer}, {Fremling}, {Pakmor},
  {Taubenberger}, {Amanullah}, {Cenko}, {Fransson}, {Goobar}, {Leloudas},
  {Taddia}, {R{\"o}pke}, {Seitenzahl}, {Sim}, \& {Sollerman}}]{Kromer16MNRAS}
{Kromer}, M., {Fremling}, C., {Pakmor}, R., {et~al.} 2016, \mnras, 459, 4428,
  \dodoi{10.1093/mnras/stw962}

\bibitem[{{Lunnan} {et~al.}(2017){Lunnan}, {Kasliwal}, {Cao}, {Hangard},
  {Yaron}, {Parrent}, {McCully}, {Gal-Yam}, {Mulchaey}, \&
  {Ben-Ami}}]{lunnan17}
{Lunnan}, R., {Kasliwal}, M.~M., {Cao}, Y., {et~al.} 2017, \apj, 836, 60,
  \dodoi{10.3847/1538-4357/836/1/60}

\bibitem[{{Lyman} {et~al.}(2014){Lyman}, {Levan}, {Church}, {Davies}, \&
  {Tanvir}}]{lyman14}
{Lyman}, J.~D., {Levan}, A.~J., {Church}, R.~P., {Davies}, M.~B., \& {Tanvir},
  N.~R. 2014, \mnras, 444, 2157, \dodoi{10.1093/mnras/stu1574}

\bibitem[{{MacLeod} {et~al.}(2014){MacLeod}, {Goldstein}, {Ramirez-Ruiz},
  {Guillochon}, \& {Samsing}}]{macleod14}
{MacLeod}, M., {Goldstein}, J., {Ramirez-Ruiz}, E., {Guillochon}, J., \&
  {Samsing}, J. 2014, \apj, 794, 9, \dodoi{10.1088/0004-637X/794/1/9}

\bibitem[{{Margalit} \& {Metzger}(2016)}]{margalit16}
{Margalit}, B., \& {Metzger}, B.~D. 2016, \mnras, 461, 1154,
  \dodoi{10.1093/mnras/stw1410}

\bibitem[{{Marquardt} {et~al.}(2015){Marquardt}, {Sim}, {Ruiter}, {Seitenzahl},
  {Ohlmann}, {Kromer}, {Pakmor}, \& {R{\"o}pke}}]{Marquardt15AA}
{Marquardt}, K.~S., {Sim}, S.~A., {Ruiter}, A.~J., {et~al.} 2015, \aap, 580,
  A118, \dodoi{10.1051/0004-6361/201525761}

\bibitem[{{Matsuoka} \& {Maeda}(2020)}]{Matsuoka20}
{Matsuoka}, T., \& {Maeda}, K. 2020, \apj, 898, 158,
  \dodoi{10.3847/1538-4357/ab9c1b}

\bibitem[{{Metzger}(2012)}]{metzger12}
{Metzger}, B.~D. 2012, \mnras, 419, 827,
  \dodoi{10.1111/j.1365-2966.2011.19747.x}

\bibitem[{{Milisavljevic} {et~al.}(2017){Milisavljevic}, {Patnaude}, {Raymond},
  {Drout}, {Margutti}, {Kamble}, {Chornock}, {Guillochon}, {Sanders},
  {Parrent}, {Lovisari}, {Chilingarian}, {Challis}, {Kirshner}, {Penny},
  {Itagaki}, {Eldridge}, \& {Moriya}}]{milisavljevic17}
{Milisavljevic}, D., {Patnaude}, D.~J., {Raymond}, J.~C., {et~al.} 2017, \apj,
  846, 50, \dodoi{10.3847/1538-4357/aa7d9f}

\bibitem[{{Moriya} \& {Eldridge}(2016)}]{moriya16}
{Moriya}, T.~J., \& {Eldridge}, J.~J. 2016, \mnras, 461, 2155,
  \dodoi{10.1093/mnras/stw1471}

\bibitem[{{Moriya} {et~al.}(2017){Moriya}, {Mazzali}, {Tominaga}, {Hachinger},
  {Blinnikov}, {Tauris}, {Takahashi}, {Tanaka}, {Langer}, \&
  {Podsiadlowski}}]{moriya17}
{Moriya}, T.~J., {Mazzali}, P.~A., {Tominaga}, N., {et~al.} 2017, \mnras, 466,
  2085, \dodoi{10.1093/mnras/stw3225}

\bibitem[{{Nakaoka} {et~al.}(2020){Nakaoka}, {Maeda}, {Yamanaka}, {Tanaka},
  {Kawabata}, {Moriya}, {Kawabata}, {Tominaga}, {Takagi}, {Imazato},
  {Morokuma}, {Sako}, {Ohsawa}, {Nagao}, {Jiang}, {Burgaz}, {Taguchi},
  {Uemura}, {Akitaya}, {Sasada}, {Isogai}, {Otsuka}, \& {Maehara}}]{Nakaoka20}
{Nakaoka}, T., {Maeda}, K., {Yamanaka}, M., {et~al.} 2020, arXiv e-prints,
  arXiv:2005.02992.
\newblock \doarXiv{2005.02992}

\bibitem[{{Pakmor} {et~al.}(2010){Pakmor}, {Kromer}, {R{\"o}pke}, {Sim},
  {Ruiter}, \& {Hillebrandt}}]{pakmor10}
{Pakmor}, R., {Kromer}, M., {R{\"o}pke}, F.~K., {et~al.} 2010, \nat, 463, 61,
  \dodoi{10.1038/nature08642}

\bibitem[{{Pakmor} {et~al.}(2012){Pakmor}, {Kromer}, {Taubenberger}, {Sim},
  {R{\"o}pke}, \& {Hillebrandt}}]{pakmor12}
{Pakmor}, R., {Kromer}, M., {Taubenberger}, S., {et~al.} 2012, \apjl, 747, L10,
  \dodoi{10.1088/2041-8205/747/1/L10}

\bibitem[{{Perets}(2014)}]{Perets2014}
{Perets}, H.~B. 2014, arXiv e-prints, arXiv:1407.2254.
\newblock \doarXiv{1407.2254}

\bibitem[{{Perets} {et~al.}(2011){Perets}, {Badenes}, {Arcavi}, {Simon}, \&
  {Gal-yam}}]{perets11}
{Perets}, H.~B., {Badenes}, C., {Arcavi}, I., {Simon}, J.~D., \& {Gal-yam}, A.
  2011, \apj, 730, 89, \dodoi{10.1088/0004-637X/730/2/89}

\bibitem[{{Perets} {et~al.}(2010){Perets}, {Gal-Yam}, {Mazzali}, {Arnett},
  {Kagan}, {Filippenko}, {Li}, {Arcavi}, {Cenko}, \& {Fox}}]{perets10}
{Perets}, H.~B., {Gal-Yam}, A., {Mazzali}, P.~A., {et~al.} 2010, \nat, 465,
  322, \dodoi{10.1038/nature09056}

\bibitem[{{Rest} {et~al.}(2005){Rest}, {Stubbs}, {Becker}, {Miknaitis},
  {Miceli}, {Covarrubias}, {Hawley}, {Smith}, {Suntzeff}, {Olsen}, {Prieto},
  {Hiriart}, {Welch}, {Cook}, {Nikolaev}, {Huber}, {Prochtor}, {Clocchiatti},
  {Minniti}, {Garg}, {Challis}, {Keller}, \& {Schmidt}}]{Rest+05}
{Rest}, A., {Stubbs}, C., {Becker}, A.~C., {et~al.} 2005, \apj, 634, 1103,
  \dodoi{10.1086/497060}

\bibitem[{{Rosswog} {et~al.}(2008){Rosswog}, {Ramirez-Ruiz}, \&
  {Hix}}]{rosswog08}
{Rosswog}, S., {Ramirez-Ruiz}, E., \& {Hix}, W.~R. 2008, \apj, 679, 1385,
  \dodoi{10.1086/528738}

\bibitem[{{Sahu} {et~al.}(2008){Sahu}, {Tanaka}, {Anupama}, {Kawabata},
  {Maeda}, {Tominaga}, {Nomoto}, {Mazzali}, \& {Prabhu}}]{sahu08}
{Sahu}, D.~K., {Tanaka}, M., {Anupama}, G.~C., {et~al.} 2008, \apj, 680, 580,
  \dodoi{10.1086/587772}

\bibitem[{{Schechter} {et~al.}(1993){Schechter}, {Mateo}, \&
  {Saha}}]{Schechter+93}
{Schechter}, P.~L., {Mateo}, M., \& {Saha}, A. 1993, \pasp, 105, 1342,
  \dodoi{10.1086/133316}

\bibitem[{{Schlafly} \& {Finkbeiner}(2011)}]{schlafly11}
{Schlafly}, E.~F., \& {Finkbeiner}, D.~P. 2011, \apj, 737, 103,
  \dodoi{10.1088/0004-637X/737/2/103}

\bibitem[{{Schlegel} {et~al.}(1998){Schlegel}, {Finkbeiner}, \&
  {Davis}}]{schlegel98}
{Schlegel}, D.~J., {Finkbeiner}, D.~P., \& {Davis}, M. 1998, \apj, 500, 525,
  \dodoi{10.1086/305772}

\bibitem[{{Seitenzahl} {et~al.}(2013{\natexlab{a}}){Seitenzahl}, {Cescutti},
  {R{\"o}pke}, {Ruiter}, \& {Pakmor}}]{seitenzahl13a}
{Seitenzahl}, I.~R., {Cescutti}, G., {R{\"o}pke}, F.~K., {Ruiter}, A.~J., \&
  {Pakmor}, R. 2013{\natexlab{a}}, \aap, 559, L5,
  \dodoi{10.1051/0004-6361/201322599}

\bibitem[{{Seitenzahl} {et~al.}(2009){Seitenzahl}, {Taubenberger}, \&
  {Sim}}]{seitenzahl2009}
{Seitenzahl}, I.~R., {Taubenberger}, S., \& {Sim}, S.~A. 2009, \mnras, 400,
  531, \dodoi{10.1111/j.1365-2966.2009.15478.x}

\bibitem[{{Seitenzahl} {et~al.}(2014){Seitenzahl}, {Timmes}, \&
  {Magkotsios}}]{seitenzahl14}
{Seitenzahl}, I.~R., {Timmes}, F.~X., \& {Magkotsios}, G. 2014, \apj, 792, 10,
  \dodoi{10.1088/0004-637X/792/1/10}

\bibitem[{{Seitenzahl} {et~al.}(2013{\natexlab{b}}){Seitenzahl},
  {Ciaraldi-Schoolmann}, {R{\"o}pke}, {Fink}, {Hillebrandt}, {Kromer},
  {Pakmor}, {Ruiter}, {Sim}, \& {Taubenberger}}]{seitenzahl13}
{Seitenzahl}, I.~R., {Ciaraldi-Schoolmann}, F., {R{\"o}pke}, F.~K., {et~al.}
  2013{\natexlab{b}}, \mnras, 429, 1156, \dodoi{10.1093/mnras/sts402}

\bibitem[{{Seitenzahl} {et~al.}(2016){Seitenzahl}, {Kromer}, {Ohlmann},
  {Ciaraldi-Schoolmann}, {Marquardt}, {Fink}, {Hillebrandt}, {Pakmor},
  {R{\"o}pke}, {Ruiter}, {Sim}, \& {Taubenberger}}]{Seitenzahl16AA}
{Seitenzahl}, I.~R., {Kromer}, M., {Ohlmann}, S.~T., {et~al.} 2016, \aap, 592,
  A57, \dodoi{10.1051/0004-6361/201527251}

\bibitem[{{Sell} {et~al.}(2015){Sell}, {Maccarone}, {Kotak}, {Knigge}, \&
  {Sand}}]{sell15}
{Sell}, P.~H., {Maccarone}, T.~J., {Kotak}, R., {Knigge}, C., \& {Sand}, D.~J.
  2015, \mnras, 450, 4198, \dodoi{10.1093/mnras/stv902}

\bibitem[{{Shappee} {et~al.}(2017){Shappee}, {Stanek}, {Kochanek}, \&
  {Garnavich}}]{shappee17}
{Shappee}, B.~J., {Stanek}, K.~Z., {Kochanek}, C.~S., \& {Garnavich}, P.~M.
  2017, \apj, 841, 48, \dodoi{10.3847/1538-4357/aa6eab}

\bibitem[{{Shen} {et~al.}(2019){Shen}, {Quataert}, \& {Pakmor}}]{shen19}
{Shen}, K.~J., {Quataert}, E., \& {Pakmor}, R. 2019, \apj, 887, 180,
  \dodoi{10.3847/1538-4357/ab5370}

\bibitem[{{Shen} \& {Schwab}(2017)}]{shen17}
{Shen}, K.~J., \& {Schwab}, J. 2017, \apj, 834, 180,
  \dodoi{10.3847/1538-4357/834/2/180}

\bibitem[{{Sim} {et~al.}(2012){Sim}, {Fink}, {Kromer}, {R{\"o}pke}, {Ruiter},
  \& {Hillebrandt}}]{Sim12MNRAS}
{Sim}, S.~A., {Fink}, M., {Kromer}, M., {et~al.} 2012, \mnras, 420, 3003,
  \dodoi{10.1111/j.1365-2966.2011.20162.x}

\bibitem[{{Sim} {et~al.}(2010){Sim}, {R{\"o}pke}, {Hillebrandt}, {Kromer},
  {Pakmor}, {Fink}, {Ruiter}, \& {Seitenzahl}}]{Sim10ApJ}
{Sim}, S.~A., {R{\"o}pke}, F.~K., {Hillebrandt}, W., {et~al.} 2010, \apjl, 714,
  L52, \dodoi{10.1088/2041-8205/714/1/L52}

\bibitem[{{Smith} {et~al.}(2010){Smith}, {Chornock}, {Silverman}, {Filippenko},
  \& {Foley}}]{smith10}
{Smith}, N., {Chornock}, R., {Silverman}, J.~M., {Filippenko}, A.~V., \&
  {Foley}, R.~J. 2010, \apj, 709, 856, \dodoi{10.1088/0004-637X/709/2/856}

\bibitem[{{Srivastav} {et~al.}(2014){Srivastav}, {Anupama}, \& {Sahu}}]{sri14}
{Srivastav}, S., {Anupama}, G.~C., \& {Sahu}, D.~K. 2014, \mnras, 445, 1932,
  \dodoi{10.1093/mnras/stu1878}

\bibitem[{{Sutherland} \& {Wheeler}(1984)}]{sutherland84}
{Sutherland}, P.~G., \& {Wheeler}, J.~C. 1984, \apj, 280, 282,
  \dodoi{10.1086/161995}

\bibitem[{{Taubenberger}(2017)}]{taubenberger17}
{Taubenberger}, S. 2017, {The Extremes of Thermonuclear Supernovae}, 317,
  \dodoi{10.1007/978-3-319-21846-5_37}

\bibitem[{{Tomita} {et~al.}(2006){Tomita}, {Deng}, {Maeda}, {Yoshii}, {Nomoto},
  {Mazzali}, {Suzuki}, {Kobayashi}, {Minezaki}, {Aoki}, {Enya}, \&
  {Suganuma}}]{tomita06}
{Tomita}, H., {Deng}, J., {Maeda}, K., {et~al.} 2006, \apj, 644, 400,
  \dodoi{10.1086/503554}

\bibitem[{{Uomoto}(1986)}]{uomoto86}
{Uomoto}, A. 1986, \apjl, 310, L35, \dodoi{10.1086/184777}

\bibitem[{{Valenti} {et~al.}(2008{\natexlab{a}}){Valenti}, {Benetti},
  {Cappellaro}, {Patat}, {Mazzali}, {Turatto}, {Hurley}, {Maeda}, {Gal-Yam},
  {Foley}, {Filippenko}, {Pastorello}, {Challis}, {Frontera}, {Harutyunyan},
  {Iye}, {Kawabata}, {Kirshner}, {Li}, {Lipkin}, {Matheson}, {Nomoto}, {Ofek},
  {Ohyama}, {Pian}, {Poznanski}, {Salvo}, {Sauer}, {Schmidt}, {Soderberg}, \&
  {Zampieri}}]{valenti08}
{Valenti}, S., {Benetti}, S., {Cappellaro}, E., {et~al.} 2008{\natexlab{a}},
  \mnras, 383, 1485, \dodoi{10.1111/j.1365-2966.2007.12647.x}

\bibitem[{{Valenti} {et~al.}(2008{\natexlab{b}}){Valenti}, {Elias-Rosa},
  {Taubenberger}, {Stanishev}, {Agnoletto}, {Sauer}, {Cappellaro},
  {Pastorello}, {Benetti}, {Riffeser}, {Hopp}, {Navasardyan}, {Tsvetkov},
  {Lorenzi}, {Patat}, {Turatto}, {Barbon}, {Ciroi}, {Di Mille}, {Frandsen},
  {Fynbo}, {Laursen}, \& {Mazzali}}]{valenti08a}
{Valenti}, S., {Elias-Rosa}, N., {Taubenberger}, S., {et~al.}
  2008{\natexlab{b}}, \apjl, 673, L155, \dodoi{10.1086/527672}

\bibitem[{{Valenti} {et~al.}(2014){Valenti}, {Yuan}, {Taubenberger}, {Maguire},
  {Pastorello}, {Benetti}, {Smartt}, {Cappellaro}, {Howell}, \&
  {Bildsten}}]{valenti14}
{Valenti}, S., {Yuan}, F., {Taubenberger}, S., {et~al.} 2014, \mnras, 437,
  1519, \dodoi{10.1093/mnras/stt1983}

\bibitem[{{Victor} {et~al.}(1994){Victor}, {Raymond}, \& {Fox}}]{victor94}
{Victor}, G.~A., {Raymond}, J.~C., \& {Fox}, J.~L. 1994, \apj, 435, 864,
  \dodoi{10.1086/174866}

\bibitem[{{Virtanen} {et~al.}(2020){Virtanen}, {Gommers}, {Oliphant},
  {Haberland}, {Reddy}, {Cournapeau}, {Burovski}, {Peterson}, {Weckesser},
  {Bright}, {van der Walt}, {Brett}, {Wilson}, {Millman}, {Mayorov}, {Nelson},
  {Jones}, {Kern}, {Larson}, {Carey}, {Polat}, {Feng}, {Moore}, {VanderPlas},
  {Laxalde}, {Perktold}, {Cimrman}, {Henriksen}, {Quintero}, {Harris},
  {Archibald}, {Ribeiro}, {Pedregosa}, {van Mulbregt}, \& {SciPy 10
  Contributors}}]{virtanen20}
{Virtanen}, P., {Gommers}, R., {Oliphant}, T.~E., {et~al.} 2020, Nature
  Methods, 17, 261, \dodoi{10.1038/s41592-019-0686-2}

\bibitem[{{Wanajo} {et~al.}(2018){Wanajo}, {M{\"u}ller}, {Janka}, \&
  {Heger}}]{wanajo18}
{Wanajo}, S., {M{\"u}ller}, B., {Janka}, H.-T., \& {Heger}, A. 2018, \apj, 852,
  40, \dodoi{10.3847/1538-4357/aa9d97}

\bibitem[{{Wheeler} {et~al.}(2015){Wheeler}, {Johnson}, \&
  {Clocchiatti}}]{wheeler15}
{Wheeler}, J.~C., {Johnson}, V., \& {Clocchiatti}, A. 2015, \mnras, 450, 1295,
  \dodoi{10.1093/mnras/stv650}

\bibitem[{{Woosley} {et~al.}(1989){Woosley}, {Pinto}, \&
  {Hartmann}}]{woosley89}
{Woosley}, S.~E., {Pinto}, P.~A., \& {Hartmann}, D. 1989, \apj, 346, 395,
  \dodoi{10.1086/168019}

\bibitem[{{Yoshida} {et~al.}(2017){Yoshida}, {Suwa}, {Umeda}, {Shibata}, \&
  {Takahashi}}]{yoshida17}
{Yoshida}, T., {Suwa}, Y., {Umeda}, H., {Shibata}, M., \& {Takahashi}, K. 2017,
  \mnras, 471, 4275, \dodoi{10.1093/mnras/stx1738}

\bibitem[{{Zenati} {et~al.}(2020){Zenati}, {Bobrick}, \& {Perets}}]{zenati20}
{Zenati}, Y., {Bobrick}, A., \& {Perets}, H.~B. 2020, \mnras, 493, 3956,
  \dodoi{10.1093/mnras/staa507}

\bibitem[{{Zenati} {et~al.}(2019{\natexlab{a}}){Zenati}, {Perets}, \&
  {Toonen}}]{zenati2019b}
{Zenati}, Y., {Perets}, H.~B., \& {Toonen}, S. 2019{\natexlab{a}}, \mnras, 486,
  1805, \dodoi{10.1093/mnras/stz316}

\bibitem[{{Zenati} {et~al.}(2019{\natexlab{b}}){Zenati}, {Toonen}, \&
  {Perets}}]{Zenati2019}
{Zenati}, Y., {Toonen}, S., \& {Perets}, H.~B. 2019{\natexlab{b}}, \mnras, 482,
  1135, \dodoi{10.1093/mnras/sty2723}

\bibitem[{{Zhang} {et~al.}(2016){Zhang}, {Wang}, {Zhang}, {Zhang},
  {Ganeshalingam}, {Li}, {Filippenko}, {Zhao}, {Zheng}, {Bai}, {Chen}, {Chen},
  {Huang}, {Mo}, {Rui}, {Song}, {Sai}, {Li}, {Wang}, \& {Wu}}]{zhang16}
{Zhang}, K., {Wang}, X., {Zhang}, J., {et~al.} 2016, \apj, 820, 67,
  \dodoi{10.3847/0004-637X/820/1/67}

\bibitem[{{Zhang} {et~al.}(2004){Zhang}, {Wang}, {Zhou}, {Li}, {Ma}, {Jiang},
  \& {Li}}]{zhang04}
{Zhang}, T., {Wang}, X., {Zhou}, X., {et~al.} 2004, \aj, 128, 1857,
  \dodoi{10.1086/423699}

\end{thebibliography}

\clearpage
\appendix

\renewcommand\thetable{A\arabic{table}} 
\setcounter{table}{0}

\begin{deluxetable*}{ccccccccc}[h!]
\tablecaption{\emph{HST} Imaging of SN~2019ehk \label{tbl:hst_table}}
\tablecolumns{9}
\tablewidth{0.45\textwidth}
\tablehead{
\colhead{Instrument} & \colhead{Aperture} & \colhead{Filter} & \colhead{MJD}& \colhead{Phase} & \colhead{Exp. Time} & \colhead{Proposal No.} & \colhead{Magnitude$^{a}$} & \colhead{Error}\\
\colhead{} & \colhead{} & \colhead{} & \colhead{}& \colhead{(days)}& \colhead{(s)} & \colhead{} & \colhead{(mag)} & \colhead{} 
}
\startdata
WFC3 & UVIS & F275W & 58877.92 & 321.78 & 2190.0 & 15654 & >26.93 & --\\
WFC3 & UVIS & F275W & 58923.58 & 321.78 & 2190.0 & 15654 & >26.59 & --\\
WFC3 & UVIS & F336W & 58877.90 & 276.10 & 1110.0 & 15654 & >26.65  & --\\
WFC3 & UVIS & F336W & 58923.58 & 321.78 & 1110.0 & 15654 & >26.55  & --\\
WFC3 & UVIS & F438W & 58877.89 & 276.10 & 1050.0 & 15654 & 25.73  & 0.10\\
WFC3 & UVIS & F438W & 58923.57 & 321.78 & 1050.0 & 15654 & >26.44  & --\\
WFC3 & UVIS & F555W & 58877.93 & 276.10 & 670.0 & 15654 & 24.38  & 0.04\\
WFC3 & UVIS & F555W & 58923.59 & 321.78 & 670.0 & 15654 & 25.26  & 0.08\\
WFC3 & UVIS & F555W & 58990.73 & 388.93 & 1500.0 & 16075 & 25.96  & 0.07\\
WFC3 & UVIS & F814W & 58877.89 & 276.10 & 836.0 & 15654 & 22.03  & 0.01\\
WFC3 & UVIS & F814W & 58923.57 & 321.78 & 836.0 & 15654 & 23.07  & 0.03\\
WFC3 & UVIS & F814W & 58990.70 & 388.93 & 900.0 & 16075 & 24.55  & 0.06\\
WFC3 & IR & F110W & 58990.64 & 388.84 & 1211.75 & 16075 &  24.88  & 0.05\\
WFC3 & IR & F160W & 58990.64 & 388.84 & 1211.75 & 16075 &  24.37  & 0.07\\
\enddata
\tablenotetext{a}{All apparent magnitudes in AB system. No extinction corrections have been made for MW or host reddening.}

\end{deluxetable*}

\begin{deluxetable}{cccc}[h!]
\tablecaption{Bolometric Light Curve Data \label{tbl:bol_table}}
\tablecolumns{4}
\tablewidth{0.45\textwidth}
\tablehead{
\colhead{MJD} & \colhead{Phase$^{a}$} & \colhead{Luminosity$^{b}$} & \colhead{Uncertainty}\\
\colhead{} & \colhead{days} & \colhead{erg s$^{-1}$} & \colhead{erg s$^{-1}$}
}
\startdata
58602.23 & +0.43 & 1.83e+41 & 1.15e+40\\
58603.17 & +1.37 & 5.94e+41 & 2.99e+40\\
58603.22 & +1.42 & 6.10e+41 & 3.07e+40\\
58603.62 & +1.82 & 6.53e+41 & 3.55e+40\\
58604.61 & +2.81 & 7.76e+41 & 4.03e+40\\
58605.25 & +3.45 & 1.75e+42 & 9.28e+40\\
58606.21 & +4.41 & 1.27e+42 & 7.12e+40\\
58606.21 & +4.41 & 1.27e+42 & 7.05e+40\\
58606.26 & +4.46 & 1.22e+42 & 6.88e+40\\
58607.24 & +5.44 & 8.80e+41 & 5.34e+40\\
58607.39 & +5.59 & 8.33e+41 & 5.25e+40\\
58607.55 & +5.75 & 7.59e+41 & 7.77e+40\\
58608.13 & +6.33 & 5.91e+41 & 3.04e+40\\
58609.18 & +7.38 & 5.48e+41 & 2.88e+40\\
58612.21 & +10.41 & 6.86e+41 & 3.75e+40\\
58612.21 & +10.41 & 6.86e+41 & 3.72e+40\\
58614.39 & +12.59 & 7.63e+41 & 4.06e+40\\
58615.14 & +13.34 & 7.84e+41 & 4.04e+40\\
58615.36 & +13.56 & 8.03e+41 & 4.27e+40\\
58616.18 & +14.38 & 8.23e+41 & 4.30e+40\\
58617.08 & +15.28 & 7.54e+41 & 4.05e+40\\
58619.19 & +17.39 & 5.87e+41 & 3.88e+40\\
58619.19 & +17.39 & 5.87e+41 & 3.81e+40\\
58622.52 & +20.72 & 4.50e+41 & 2.88e+40\\
58626.26 & +24.46 & 3.49e+41 & 1.93e+40\\
58628.30 & +26.50 & 3.21e+41 & 1.80e+40\\
58631.12 & +29.32 & 2.81e+41 & 1.62e+40\\
58632.18 & +30.38 & 2.74e+41 & 2.02e+40\\
58633.20 & +31.40 & 2.72e+41 & 1.62e+40\\
\enddata
\tablenotetext{a}{Relative to explosion.}
\tablenotetext{b}{Covers wavelength range 3000-10000\,\AA}
\end{deluxetable}

\begin{deluxetable}{cccc}[h!]
\tablecaption{Bolometric Light Curve Data  (Cont.)\label{tbl:bol_table2}}
\tablecolumns{4}
\tablewidth{0.45\textwidth}
\tablehead{
\colhead{MJD} & \colhead{Phase$^{a}$} & \colhead{Luminosity$^{b}$} & \colhead{Uncertainty}\\
\colhead{} & \colhead{days} & \colhead{erg s$^{-1}$} & \colhead{erg s$^{-1}$}
}
\startdata
58633.20 & +31.40 & 2.72e+41 & 1.64e+40\\
58633.27 & +31.47 & 2.66e+41 & 1.52e+40\\
58634.18 & +32.38 & 2.58e+41 & 1.49e+40\\
58636.08 & +34.28 & 2.45e+41 & 1.34e+40\\
58636.11 & +34.31 & 2.45e+41 & 1.34e+40\\
58636.21 & +34.41 & 2.44e+41 & 1.46e+40\\
58636.35 & +34.55 & 2.47e+41 & 1.51e+40\\
58639.05 & +37.25 & 2.17e+41 & 1.29e+40\\
58639.18 & +37.38 & 2.16e+41 & 1.34e+40\\
58640.18 & +38.38 & 2.10e+41 & 1.30e+40\\
58642.10 & +40.30 & 1.99e+41 & 1.26e+40\\
58642.22 & +40.42 & 2.00e+41 & 1.31e+40\\
58644.07 & +42.27 & 2.05e+41 & 1.25e+40\\
58646.23 & +44.43 & 1.95e+41 & 1.43e+40\\
58649.22 & +47.42 & 1.87e+41 & 1.40e+40\\
58652.28 & +50.48 & 1.70e+41 & 1.49e+40\\
58652.71 & +50.91 & 1.70e+41 & 1.34e+40\\
58657.53 & +55.73 & 1.52e+41 & 9.97e+39\\
58658.04 & +56.24 & 1.48e+41 & 8.88e+39\\
58658.18 & +56.38 & 1.46e+41 & 9.21e+39\\
58661.20 & +59.40 & 1.32e+41 & 8.76e+39\\
58670.01 & +68.21 & 9.83e+40 & 6.95e+39\\
58687.86 & +86.06 & 6.99e+40 & 6.94e+39\\
58688.97 & +87.17 & 6.64e+40 & 5.75e+39\\
58690.97 & +89.17 & 6.75e+40 & 5.33e+39\\
58696.97 & +95.17 & 5.39e+40 & 4.26e+39\\
58877.93 & +276.13 & 1.07e+39 & 4.68e+37\\
58923.59 & +321.79 & 4.76e+38 & 2.19e+37\\
58990.73 & +388.93 & 1.55e+38 & 9.05e+36\\
\enddata
\tablenotetext{a}{Relative to explosion.}
\tablenotetext{b}{Covers wavelength range 3000-10000\,\AA}
\end{deluxetable}

\begin{deluxetable*}{lcccccc}
\centering
\tablecaption{Explosion Model Characteristics\label{tab:explosion_models}}
\tablecolumns{7}
\tablewidth{0.99\textwidth}

\tablehead{
\colhead{Model} &
\colhead{Description} & \colhead{SN Type} & \colhead{$M_{ej}$} & \colhead{M($^{56}$Ni)}& 
\colhead{$^{57}\textrm{Co}/^{56}\textrm{Co}$} & \colhead{Reference}\\
& & & $(\textrm{M}_{\odot})$ &  $(\textrm{M}_{\odot})$ & &}
\startdata
W7 &  Deflagration$^a$ & SN~Ia & $1.38$ & $0.59$ & 0.041 & \citealt{iwamoto99} \\
ddt$\_$n100 & Delayed Detonation$^a$ & SN~Ia & 1.40 & 0.60 & 0.031 & \citealt{seitenzahl13} \\
det$\_$1.06 & Detonation$^a$ & SN~Ia & $1.06$ & $0.56$ & 0.006 & \citealt{Sim10ApJ} \\
doubledt$\_$CSDD-S & Double Detonation$^a$ & SN~Ia & $0.79$ & $0.21$ & 0.044 & \citealt{Sim12MNRAS} \\
def$\_$N100def & Pure Deflagration$^a$ & SN~Ia & $1.40$ & $0.36$ & 0.038 & \citealt{Fink14MNRAS} \\
det$\_$ONe15e7 & O-Ne WD Detonation$^a$ & SN~Ia & $1.23$ & $0.96$ & 0.009 & \citealt{Marquardt15AA} \\
gcd$\_$GCD200 & Detonation$^a$ & SN~Ia & $1.40$ & $0.74$ & 0.025 & \citealt{Seitenzahl16AA}\\
merger$\_$11+09 & Violent Merger$^b$ & SN~Ia & $1.95$ & $0.10$ & 0.024 & \citealt{pakmor12} \\
merger$\_$09+09 & Violent Merger$^b$ & SN~Ia & $1.75$ & $0.10$ & 0.003 & \citealt{pakmor10} \\
merger$\_$09+076$\_$Z1 & Violent Merger$^b$ & SN~Ia & $1.6$ & $0.18$ & 0.009 & \citealt{Kromer13ApJ} \\
merger$\_$09+076$\_$Z0.01 & Violent Merger$^b$ & SN~Ia & $1.6$ & $0.18$ & 0.003 & \citealt{Kromer16MNRAS} \\
0.55+0.63$\_$Carich & WD Merger & \ca & $0.45$ & $0.013$ & 0.0028 & Zenati et al. 2020b, in prep. \\
0.52+0.63$\_$Carich & WD Merger & \ca & $0.43$ & $0.052$ & 0.00084 & Zenati et al. 2020b, in prep. \\
0.50+0.58$\_$Carich & WD Merger & \ca & $0.36$ & $0.011$ & 0.011 & Zenati et al. 2020b, in prep. \\
03HeWD+14NS & NS + He WD & FRRT$^c$ & $0.30$ & $0.0036$ & $0.049$ & Private Communication (A. Bobrick)\\
063COWD+14NS & NS + CO WD & FRRT$^c$ & $0.63$ & $0.0049$ & 0.040 & \citealt{zenati20} \\
063COWD+20NS & NS + CO WD & FRRT$^c$ & $0.63$ & $0.0061$ & 0.058 & \citealt{zenati20} \\
08COWD+14NS & NS + CO WD & FRRT$^c$ & $0.80$ & $0.029$ & 0.078 & \citealt{zenati20} \\
09ONeWD+14NS & NS + ONe WD & FRRT$^c$ & $0.9$ & $0.023$ & $0.120$ & Bobrick et al. 2020b, in prep. \\
09COWD14NS & NS + CO WD & FRRT$^c$ & $0.9$ & $0.026$ & 0.11 & Bobrick et al. 2020b, in prep. \\
10ONeWD14NS & NS + ONe WD & FRRT$^c$ & $0.9$ & $0.029$ & $0.11$ & Bobrick et al. 2020b, in prep. \\
11ONeWD14NS & NS + ONe WD & FRRT$^c$ & $1.1$ & $0.046$ & $0.093$ & Bobrick et al. 2020b, in prep. \\
12ONeWD14NS & NS + ONe WD & FRRT$^c$ & $1.2$ & $0.054$ & $0.11$ & Bobrick et al. 2020b, in prep. \\
12ONeWD20NS & NS + ONe WD & FRRT$^c$ & $1.2$ & $0.034$ & $0.068$ & Bobrick et al. 2020b, in prep. \\
12ONeWD50BH & BH + ONe WD & FRRT$^c$ & $1.2$ & $0.010$ & $0.044$ & Bobrick et al. 2020b, in prep. \\
13ONeWD14NS & BH + ONe WD & FRRT$^c$ & $1.3$ & $0.090$ & $0.072$ & Bobrick et al. 2020b, in prep. \\
CO145 & CO Star Core-Collapse & USSN$^d$ & $0.098$ & $0.0097$ & 0.046 & \citealt{yoshida17} \\
CO150 & CO Star Core-Collapse & USSN$^d$ & $0.11$ & $0.0057$ & 0.041 & \citealt{yoshida17} \\
ussn$\_$E1e50erg & Core-Collapse, $10^{50}$~erg & USSN$^d$ & $0.20$ & $0.026$ & 0.091 & \citealt{moriya17} \\
ussn$\_$E2.5e50erg & Core-Collapse, $2.5 \times 10^{50}$~erg & USSN$^d$ & $0.20$ & $0.030$ & 0.085 & \citealt{moriya17} \\
ussn$\_$E5e50erg & Core-Collapse, $5 \times 10^{50}$~erg & USSN$^d$ & $0.20$ & $0.034$ & 0.080 & \citealt{moriya17} \\
e8.8 & Core-Collapse, $M_{\rm ZAMS}$ = 8.8~$\Msun$ & ECSN$^e$ & $0.017$ & $0.0029$ & 0.034 & \citealt{wanajo18} \\
z9.6 & Core-Collapse, $M_{\rm ZAMS}$ = 9.6~$\Msun$ & CCSN$^f$ & $0.56$ & $0.0025$ & 0.036 & \citealt{wanajo18} \\
u8.1 & Core-Collapse, $M_{\rm ZAMS}$ = 8.1~$\Msun$ & CCSN$^f$ & $0.33$ & $0.0016$ & 0.046 & \citealt{wanajo18} \\
s11 & Core-Collapse, $M_{\rm ZAMS}$ = 11~$\Msun$ & CCSN$^f$ & $1.48$ & $0.0039$ & 0.023 & \citealt{wanajo18} \\
\enddata
\tablenotetext{a}{Single Degenerate Channel}
\tablenotetext{b}{Double Degenerate Channel}
\tablenotetext{c}{``Faint Rapid Red Transient''}
\tablenotetext{d}{``Ultra-stripped supernova''}
\tablenotetext{e}{``Electron-capture supernova''}
\tablenotetext{f}{``Core-collapse supernova''}
\end{deluxetable*}


\end{document}